\documentclass[preprint,12pt]{elsarticle}

\usepackage{amssymb}

\usepackage[utf8]{inputenc}
\usepackage{hyperref}
\usepackage[nottoc,numbib]{tocbibind}
\usepackage{setspace}
\usepackage{booktabs}

\usepackage{listings}
\usepackage{xcolor}

\usepackage{caption}
\usepackage{subcaption}

\usepackage{amsmath}

\definecolor{codegreen}{rgb}{0,0.6,0}
\definecolor{codegray}{rgb}{0.5,0.5,0.5}
\definecolor{codepurple}{rgb}{0.58,0,0.82}
\definecolor{backcolour}{rgb}{0.95,0.95,0.92}

\lstdefinestyle{mystyle}{
    backgroundcolor=\color{backcolour},   
    commentstyle=\color{codegreen},
    keywordstyle=\color{blue},
    numberstyle=\tiny\color{codegray},
    stringstyle=\color{codepurple},
    basicstyle=\ttfamily\footnotesize,
    breakatwhitespace=false,         
    breaklines=true,                 
    captionpos=b,                    
    keepspaces=true,                 
    numbers=left,                    
    numbersep=5pt,                  
    showspaces=false,                
    showstringspaces=false,
    showtabs=false,                  
    tabsize=2
}
\lstdefinelanguage{Ini}{
    basicstyle=\ttfamily\small,
    columns=fullflexible,
    morecomment=[s][\color{codegreen}\bfseries]{[}{]},
    morecomment=[l]{\#},
    morecomment=[l]{;},
    commentstyle=\color{gray}\ttfamily,
    morekeywords={},
    otherkeywords={=,:},
    keywordstyle={\color{green}\bfseries}
}

\journal{Comp. Phys. Comm.}

\begin{document}

\begin{frontmatter}
\title{Hermes-3: Multi-component plasma simulations with BOUT++}

\author[llnl]{Ben Dudson}
\author[york]{Mike Kryjak}
\author[york,ccfe]{Hasan Muhammed}
\author[york]{Peter Hill}
\author[ccfe]{John Omotani}

\affiliation[llnl]{organization={Lawrence Livermore National Laboratory},
            addressline={7000 East Avenue},
            city={Livermore},
            postcode={94550},
            state={CA},
            country={USA}}

\affiliation[york]{organization={School of Physics, Engineering and Technology, University of York},
            addressline={Heslington},
            city={York},
            postcode={YO10 5DD},
            country={UK}}

\affiliation[ccfe]{organization={United Kingdom Atomic Energy Authority},
            addressline={Culham Centre for Fusion Energy, Culham Science Centre},
            city={Abingdon},
            postcode={OX14 3DB},
            country={UK}}

\begin{abstract}
  A new open source tool for fluid simulation of multi-component
  plasmas is presented, based on a flexible software design that is
  applicable to scientific simulations in a wide range of fields. This
  design enables the same code to be configured at run-time to solve
  systems of partial differential equations in 1D, 2D or 3D, either
  for transport (steady-state) or turbulent (time-evolving) problems,
  with an arbitrary number of ion and neutral species.

  To demonstrate the capabilities of this tool, applications relevant
  to the boundary of tokamak plasmas are presented: 1D simulations of
  diveror plasmas evolving equations for all charge states of neon and
  deuterium; 2D transport simulations of tokamak equilibria in
  single-null X-point geometry with plasma ion and neutral atom
  species; and simulations of the time-dependent propagation of plasma
  filaments (blobs).

  Hermes-3 is publicly available on Github under the GPL-3 open source
  license.  The repository includes documentation and a suite of unit,
  integrated and convergence tests.
\end{abstract}


\end{frontmatter}

{\bf PROGRAM SUMMARY}

\begin{small}
\noindent
{\em Program Title:} Hermes-3  \\
{\em CPC Library link to program files:}  \\
{\em Developer's repository link:} https://github.com/bendudson/hermes-3 \\
{\em Code Ocean capsule:} \\
{\em Licensing provisions:} GPLv3 \\
{\em Programming language:} C++14 \\
{\em Nature of problem:}
Simulation of dynamics and steady-state solutions of multiple ion and neutral species
in magnetised plasmas using drift-reduced fluid plasma models in 1 to 3 spatial dimensions. 
Focused on but not restricted to the simulation of turbulence in the boundary of tokamaks. \\
{\em Solution method:}
Hermes-3 implements a system of flexible software components, that are configured
at run-time and coupled by passing a state object between them (the Encapsulate State
and Command patterns). Hermes-3 builds on BOUT++ [1], employing the Method of Lines
with implicit and explicit time-integration methods in curvilinear coordinates on block-structured
meshes, making use of libraries including SUNDIALS, PETSc and Hypre.
Hermes-3 implements drift-reduced plasma fluid equations using conservative finite difference methods,
and atomic processes that couple plasma and neutral fluids.

\end{small}

\section{Introduction}
\label{sec:introduction}

An important feature of magnetically confined fusion plasmas in
tokamaks is that they consist of a mixture of different ion
species. Computational studies of fusion plasma phenomena, such as
plasma turbulence, have to date frequently treated plasmas as if they
consisted of a single ion species, typically deuterium due to its use
in experiments. In order to make predictions of phenomena important to
the performance and design of the divertor of future fusion reactors,
models must capture the interactions between plasma and neutral gas
species (deuterium and tritium in a reactor), as well as radiation
from impurity species (eg. Be, C, Ne, Ar, W), and pumping of helium
``ash'' and other species.  In many cases none of these species can be
treated as ``trace'' but are all coupled, so that all should be
considered self-consistently in the same simulation. For each of these
atomic species the density and dynamics of multiple states may need
to be considered, for example the ground state, multiple ionisation
stages, and molecular species. In some cases it may also be necessary
to track metastable states, such as vibrationally excited states or
charged molecules. The result is a potentially large number of species
types which must be solved for, together with a complex set of
reactions between them.

Historically there has been a division in terms of simulation tools
and to an extent also research communities, between multi-fluid
transport codes such as SOLPS~\cite{schneider-2006},
UEDGE~\cite{rognlien-2002}, EDGE2D~\cite{simonini-1994} and
BOUT++/trans-neut~\cite{wang2014}, which employ simplified models for
the cross-field transport (typically diffusive) but evolve many
different species, and the turbulence codes including
GBS~\cite{giacomin2021,halpern2016,ricci2012},
TOKAM3X~\cite{tamain2010,tamain2016}, (H)ESEL~\cite{madsen2016a}, and
various models built on
BOUT++~\cite{Dudson2009,dudson2015,bout:manual} such as
Hermes~\cite{Dudson2017} and STORM~\cite{easy2014,ukaea:storm}. These
latter models can solve for the 3D time-varying turbulent transport,
but typically only evolve a single ion species. During the last 5
years, and currently ongoing, are several efforts
(e.g.~\cite{BUFFERAND201982, coroado2021, shrish2021}) to develop codes which can solve for the
turbulent transport self consistently with multiple ion and neutral
gas species~\cite{LEDDY2017994}.

If the number of species states to be solved for is relatively small,
such as electrons, ions and a single atomic species as in
Hermes-2~\cite{dudson:hermes2} and recent versions of
STORM~\cite{ukaea:storm}, then code can be written for each
species. Unfortunately with this design the size of the code (number
of lines) grows linearly with the number of species states, becoming
increasingly error prone, difficult to test, and hard to maintain, as
the number of species is increased. The authors' experience with
Hermes-2 indicated that this was not a viable path to multi-species
simulations.

Here an open-source multi-fluid plasma simulation tool is presented,
along with the flexible software design which makes the resulting
model and software complexity manageable.  It is built on BOUT++,
which provides low-level data management and operations, and augments
this with a reusable model component system. By doing this we unify 1D
tokamak divertor simulation code SD1D~\cite{dudson2019} with 2D and 3D
transport and turbulence code Hermes~\cite{Dudson2017}, and enable
these tools to be extended towards multiple ion species simulations
that self-consistently include both plasma turbulence and transport
(atomic) physics.

In section~\ref{sec:numerical} the numerical methods are described; in
section~\ref{sec:software} the software architecture is outlined and
compared to prior work; and in section~\ref{sec:applications} a series
of applications are presented, which demonstrate some of the
capabilities of the new code. We conclude in section~\ref{sec:conclusions}.

\section{Numerical methods}
\label{sec:numerical}

In this section we describe the numerical methods used in Hermes-3
components. The architecture described in
section~\ref{sec:software-arch} does not enforce the choice of
numerical method, but this section describes the methods that have
been implemented to date and are used in
section~\ref{sec:applications}.  The system of PDEs is solved using
the method of lines, in which the time and spatial dimensions are
treated separately: The time integrator simply integrates a set of
Ordinary Differential Equations (ODEs), and is discussed in
section~\ref{sec:time-integration}; The spatial discretisation is by
finite difference methods
(section~\ref{sec:finite_differencing}). Boundary conditions are
discussed in section~\ref{sec:boundary}.

\subsection{Time integration}
\label{sec:time-integration}

Hermes-3 is built on the BOUT++ framework~\cite{bout:manual}, and so
can make use of a range of explicit (e.g RK4), fully implicit
(e.g. BDF via SUNDIALS~\cite{hindmarsh2005}), and implicit-explicit
(e.g IMEX-BDF2 and ARKODE via SUNDIALS) methods. These were
implemented with the aim of studying time-dependent problems, such as
the study of Edge Localised Mode (ELM) eruptions~\cite{xu2010} or
tokamak edge turbulence~\cite{seto2019}, where accurate time evolution
is required.

Many of the problems of interest for Hermes-3 are steady state:
Axisymmetric tokamak transport solutions, potentially as a starting
point for 3D time-dependent turbulence simulations. To find these
steady-state solutions efficiently it is desirable to take large
timesteps, ideally an infinite timestep, damping transient
oscillations in the system. A dissipative time integration scheme that
is unconditionally stable is therefore desirable. A first order
backward Euler method and preconditioning algorithm similar to that
used in UEDGE~\cite{rognlien-2002} has been implemented here. This
method is stable for any timestep provided that the nonlinear solver,
typically a variety of Newton's method, iterations converge. In
practice this limits the timestep to a finite, and sometimes quite
small, value.

An important ingredient to robustly and efficiently solving the
nonlinear problem at each time step is a good preconditioner: It
enables the linear inner solve to converge with fewer iterations (and
so computational cost) for larger timesteps than would otherwise be
possible. Custom preconditioners developed using a methodology such as
physics-based preconditioning~\cite{chacon2002, chacon2008} can be
highly effective, but are challenging in multi-fluid contexts: The
tokamak edge is a highly nonlinear system, with a potentially large
number of species (and so equations), coupled through atomic rates
which are typically tabulated rather than analytic, and which vary by
orders of magnitude over relatively small temperature ranges. The
approach used in UEDGE, and adopted for the Backward Euler solver
here, is to use finite differences to calculate the elements of the
system Jacobian $\mathbb{J}_{ij}=\frac{\partial}{\partial
  f_j}\frac{\partial f_i}{\partial t}$ where $\mathbf{f}$ is the
vector of evolving quantites. This approximate Jacobian is used to
construct a preconditioner that (approximately) solves the linear
operator $\mathbb{I} - \Delta t\mathbb{J}$ for timestep $\Delta
t$. Methods based on incomplete LU factorization (ILU) have been
found to be effective: In serial the PETSc ILU(k) solver is used, and
in parallel the Euler library~\cite{hysom2001} within
{\it hypre}~\cite{falgout2006}, both with default settings.

The calculation of a dense Jacobian using finite differences would be
prohibitively slow in most cases: A typical simulation might contain
$N \simeq 10^5-10^8$ evolving quantities, while the Jacobian has $N^2$
elements. Fortunately the Jacobian is typically sparse, because the
finite differences and other interaction terms are local. This is
exploited by using the PETSc coloring
facilities~\cite{petsc-user-ref}, which are provided with the matrix
structure (determined by the finite difference stencil), and
efficiently calculate many Jacobian matrix entries
simultaneously. Because each cell only has a fixed number of
neighbours, the cost of evaluating the Jacobian is reduced from
scaling like $N^2$ to scaling approximately linearly with the number
of grid cells $N$.

The effectiveness of this time integrator for steady state problems
will be applied to 1D transport problems in
section~\ref{sec:1d-transport} (fig~\ref{fig:timederivs-rhsevals}),
and for 2D transport in an axisymmetric tokamak geometry in
section~\ref{sec:applications-2d}.

\subsection{Finite differencing spatial operators}
\label{sec:finite_differencing}

The models to be shown here use conservative finite difference
operators that were implemented in the Hermes code~\cite{Dudson2017}
and have been improved over time and moved into the BOUT++
library. All quantities are cell centred, and advection operators are
written in terms of fluxes between cells calculated at cell faces. The
cross-field operators presently assume that the grid is orthogonal in
the tokamak poloidal plane. This limits the accuracy with which
strongly shaped divertor geometries can be simulated with the present
code. Non-orthogonal grids which align with wall surfaces can be
generated for Hermes using the BOUT++ grid generator~\cite{hypnotoad},
but the required off-diagonal metric terms have not yet been
implemented. Those terms have long been implemented in UEDGE, and were
recently added to SOLPS~\cite{dekeyser2019}, where they were found to
be essential for fluid neutral modelling on distorted grids, but
relatively unimportant when kinetic neutrals were used. Implementing
these terms is a high priority for future improvements to Hermes-3.

Because all quantities are cell centred, in the absence of dissipation
zig-zag modes are likely to develop. In~\cite{Dudson2017} an Added
Dissipation~\cite{murthy-2002} artificial dissipation term was used in
advection operators along the magnetic field.  Here this is replaced
with an HLL type flux splitting method~\cite{harten1983, donat-1996},
that improves on methods used previously for 1D tokamak divertor
simulations~\cite{dudson2019,bout:sd1d}. Results shown here use either
the MinMod or Monotonised Central (MC) slope
limiters~\cite{vanleer1977}, that have been found to provide a good
balance of numerical dissipation and stability. Details of the method
are given in \ref{apx:numerics1d}.

\begin{figure}[h]
  \centering
  \includegraphics[width=0.9\textwidth]{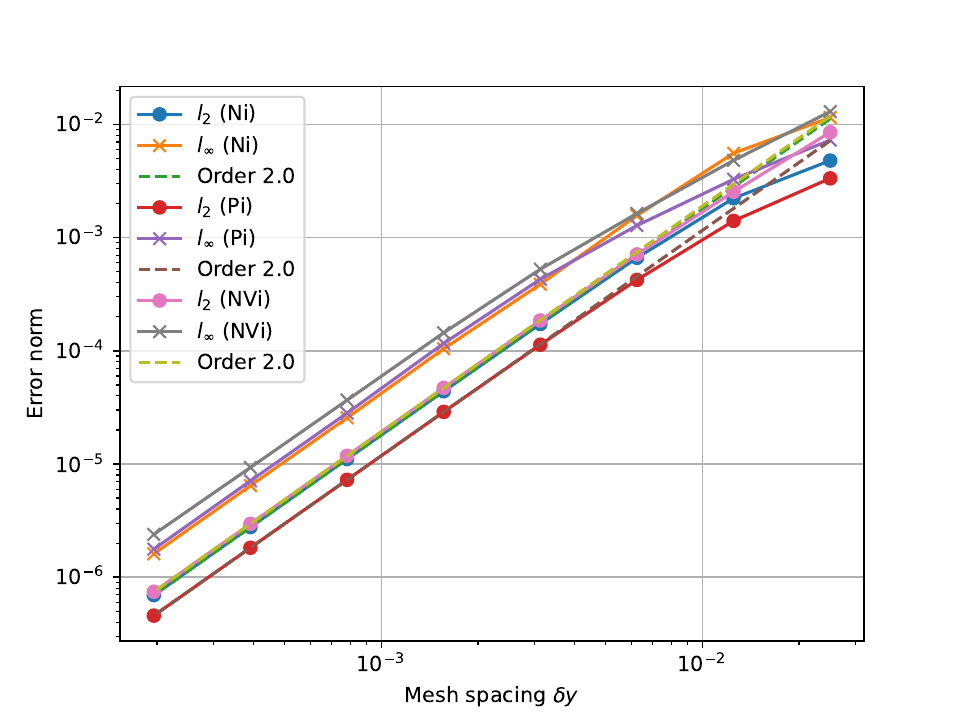}
  \caption{Verification of the convergence of a 1D system of fluid
    equations on a periodic domain.  Showing $l_2$ (Root-Mean-Square)
    and $l_\infty$ (Max) errors for the evolving density $N_i$,
    pressure $P_i$ and momentum $NV_i$. Figure produced by
    \texttt{tests/integrated/1D-fluid} in the Hermes-3
    repository~\cite{dudson:hermes3}.}
  \label{fig:fluid-norm}
\end{figure}

To verify the implementation of fluid flow along the magnetic field
for smooth solutions, a set of 1D fluid equations along a magnetic
field given in equation~\ref{eq:fluid-equations} is tested using the
Method of Manufactured solutions (MMS). This testing method has become
widely used to verify the correct implementation of complex sets of
equations, in tokamak edge plasma codes~\cite{riva2014} including
BOUT++~\cite{dudson-2015}.
\begin{subequations}
  \label{eq:fluid-equations}
  \begin{align}
    \frac{\partial n}{\partial t} &= -\nabla\cdot\left(n\mathbf{b}v_{||}\right) \\
    \frac{\partial p}{\partial t} &= -\nabla\cdot\left(p\mathbf{b}v_{||}\right) - \frac{2}{3}p\nabla\cdot\left(\mathbf{b}v_{||}\right) \\
    \frac{\partial}{\partial t}\left(mnv_{||}\right) &= -\nabla\cdot\left(mnv_{||}\mathbf{b}v_{||}\right) - \partial_{||}p
  \end{align}
\end{subequations}
Error norms as a function of mesh cell spacing are presented in
figure~\ref{fig:fluid-norm}, showing convergence towards the
manufactured solution on a 1D periodic domain. Second order convergence
is found for both $l_2$ and $l_\infty$ error norms, consistent with the
expected order of accuracy of the numerical methods.

\begin{figure}[h]
  \centering
  \includegraphics[width=0.49\textwidth]{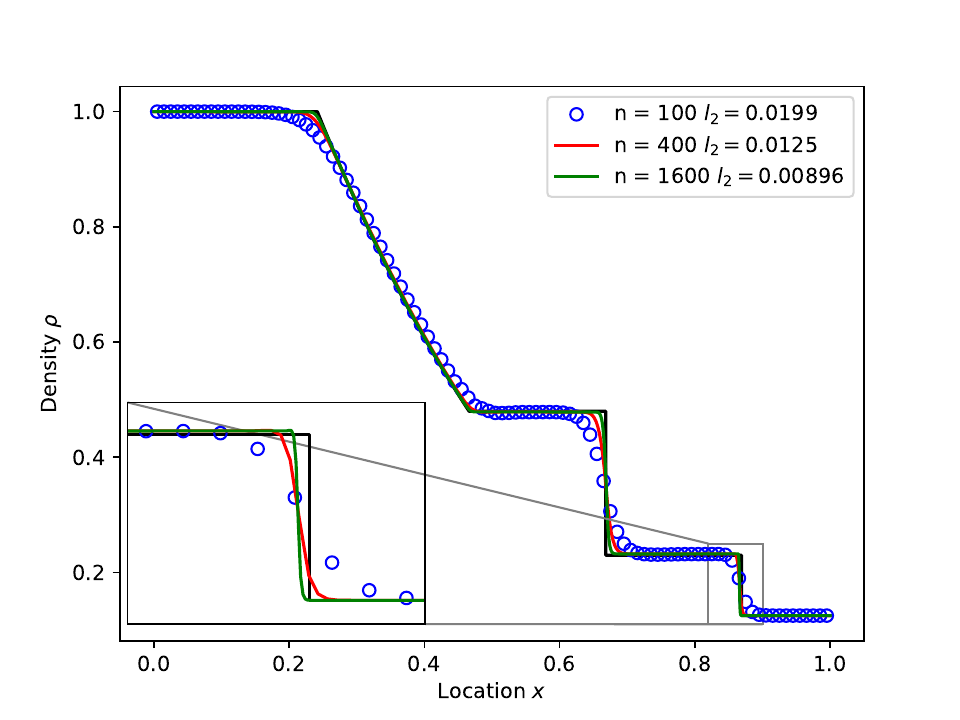}
  \includegraphics[width=0.49\textwidth]{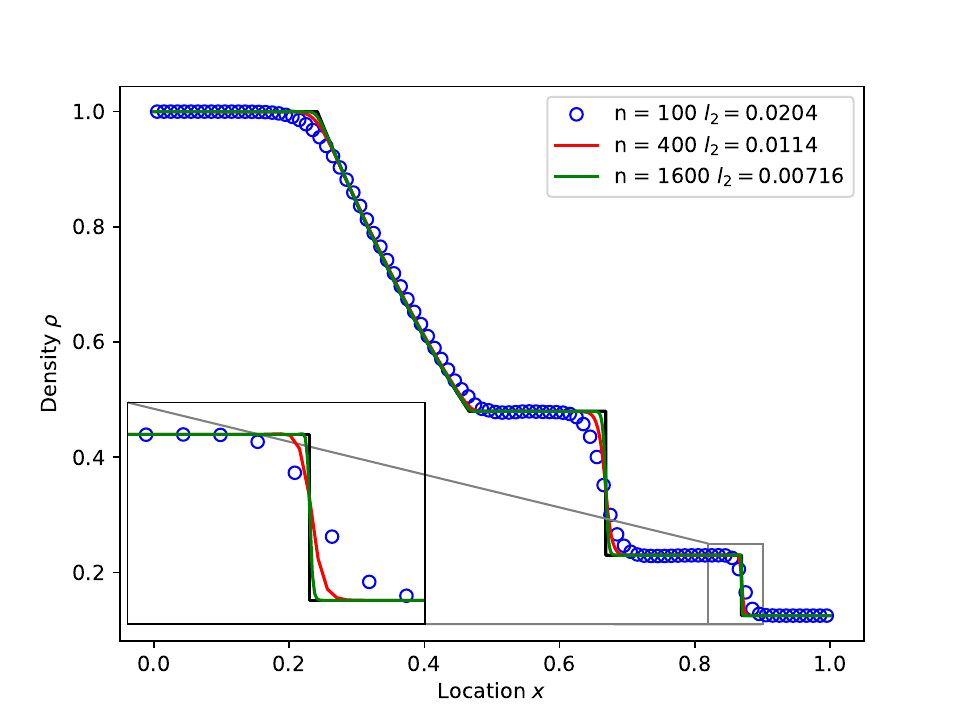}
  \caption{Standard Sod shock tube problem~\cite{sod1978} at $t=0.2$,
    with exact (analytic) solution in solid black. A solution with
    reference resolution ($n = 100$ cells) is compared to higher
    resolutions and their $l_2$ (Root-Mean-Square) errors. {\bf Left}: Solving the
    pressure form of the fluid equations. {\bf Right}: Solving the
    energy form. The inset figures show the shock front in more
    detail. Figure produced by \texttt{tests/integrated/sod-shock} and \texttt{tests/integrated/sod-shock-energy}.}
  \label{fig:sod-shock}
\end{figure}

The intended application of Hermes-3 is to magnetically confined
fusion plasmas, in which flows are typically subsonic. Nevertheless
the code must be robust to transients, and transitions to supersonic
flow can occur in tokamak plasmas~\cite{ghendrih2011}. Figure~\ref{fig:sod-shock}
shows the results of the standard 1D Sod shock tube test case~\cite{sod1978}.
In the expanded view shown inset in the left
figure~\ref{fig:sod-shock} it can be seen that the numerical shock
location lags the exact solution, so that the $l^2$ error norm does
not converge to zero. The result is insensitive to time integration
method, being observed with both the default CVODE time integrator and
the RK3-SSP method implemented in BOUT++. This is due to the choice of
pressure as an evolving quantity in equation~\ref{eq:fluid-equations},
and so pressure fluxes are calculated at cell edges.

Hermes-3 is structured in a modular way that allows equations to be
changed in the input file. If rather than evolving pressure we choose
to evolve energy density $\mathcal{E} = \frac{3}{2}p +
\frac{1}{2}mnv_{||}^2$, then the result is shown on the right of
figure~\ref{fig:sod-shock}.  Solving the fluid equations in this
conservative form recovers the correct shock front location (shown
inset in figure~\ref{fig:sod-shock} right). Further variations on this
1D shock tube problem~\cite{toro09} are included in the Hermes-3 test
suite.

We conclude that the methods currently implemented are $2^{nd}$-order
accurate for smooth solutions (figure~\ref{fig:fluid-norm}), and
remain robust around shocks, contact discontinuities and expansion
waves (figure~\ref{fig:sod-shock}).  The modular nature of Hermes-3
allows multiple fluid formulations to be implemented and inter-operate
to suit different applications.

\subsection{Boundary conditions}
\label{sec:boundary}

The domain typically solved for in 2D and 3D Hermes-3 tokamak
simulations is an annulus consisting of a region of closed and open
magnetic flux surfaces.  An example is discussed in
section~\ref{sec:applications-2d} and shown in
figure~\ref{fig:2d_domain}. The hot ``core'' of the plasma is not
modelled because the fluid equations solved become invalid in that
region. Instead a boundary condition must be imposed at that innermost
surface where no boundary physically exists. At the outer edge of the
domain the grid is typically close to, but not aligned with, the solid
vacuum vessel of the tokamak. Boundary conditions for the
thermodynamic variables on both ``core'' and ``wall'' boundaries are
typically set to either Dirichlet or Neumann.

The boundary condition on the potential $\phi$ is a variation on the
method used in the STORM model~\cite{ukaea:storm, easy2014}: A
time-evolving boundary condition that relaxes towards a Neumann
boundary. This is implemented in the following way: When inverting the
Laplacian-type equation for $\phi$ from vorticity, the potential is
fixed at both core and wall boundaries. If a simple Dirichlet
condition is used then narrow boundary layers typically form close to
the boundaries in which the imposed boundary potential is matched to
the plasma potential. These boundary layers can develop unphysical
instabilities. Instead, at every timestep the value of the boundary
condition is adjusted towards the value inside the domain with a
characteristic timescale that is set by default to $1\mu s$. In this
manner the electrostatic potential $\phi$ evolves smoothly to
solutions that can have different potentials on core and wall
boundaries.

\section{Software architecture}
\label{sec:software}

Hermes-3~\cite{dudson:hermes3, dudson:hermes3-manual} aims to
support a wide range of different models, with an arbitrary number of
species and equations. This flexibility presents a challenge for the
software design: A poorly chosen architecture will result in the code
complexity growing rapidly with the model size, so that further
progress becomes increasingly difficult as the model is extended.

There are many domains besides tokamak plasma physics where
performance is important, and where many different software components
have to interact in complex ways which need to be extended over time
as the software grows and is applied to new problems. A variety of
approaches have been developed within scientific computing and in
other fields. In section~\ref{sec:software-survey} we briefly describe
some of the approaches that influenced the design of Hermes-3.

\subsection{A brief survey of approaches}
\label{sec:software-survey}

Within physical sciences there are a number of codes which have
adopted designs that enable users and developers to develop components
or plugins, and to combine them in novel ways so that the ecosystem
becomes increasingly useful as new components are added. An example is
LAMMPS~\cite{lammps2022}, which uses a system of ``styles'' that
define interfaces which users can implement to modify the simulation
behaviour.

In the software industry Entity Component Systems (ECS) are a design
pattern which is commonly used in game development. Those are intended
to describe a set of ``Entities'' that have defined sets of behaviour
and can interact with each other. This design pattern offers
flexibility through composition rather than inheritance, and
considerable run-time configurability. A widely used and
high-performance implementation of an ECS is EnTT~\cite{entt}.

Task graphs are another widely applicable and powerful approach to
thinking about computations, which focuses on managing the
dependencies between components, so that at a high level the whole
calculation is a directed acyclic graph (DAG). Examples of task-based
systems include StarPU~\cite{starpu} and TaskFlow\cite{taskflow}.

An important aspect of all of these approaches is splitting complex
models into simpler components, which interact through standardised
interfaces and not global state.  This facilitates testing, in
particular unit testing, provides a powerful way to mitigate the
growth of complexity, and helps to maintain productivity as code
becomes larger.

\subsection{The design of Hermes-3}
\label{sec:software-arch}

The design of Hermes-3 is a combination of the Encapsulate
Context~\cite{kelly2003} and Command patterns~\cite{gof1994}.  The
main elements are a flexible store or database, into which values
(e.g. spatially dependent fields like densities, temperatures) can be
inserted and later retrieved; and a collection of composable model
components that set and use values in the store. The approach has
similarities to data oriented design~\cite{joshi2007}, in which loose
coupling between components is achieved by focusing on defining the
data being operated on. In fusion an example of this is the OMFIT
framework~\cite{Meneghini_2015}, which uses a tree data structure to
loosely couple data sources, codes and analysis scripts.

The data in a simulation is physical quantities, such as density and
temperature fields, and derived quantities representing terms in the
equations being solved. Hermes-3 stores these quantities in a nested
dictionary structure (a tree), using C++ \texttt{variant} to enable
different data types to be stored. A schema defines a convention for
where values are stored, for example
\texttt{state["species"]["h+"]["density"]} is the number density of
hydrogen ions.

Operations on the simulation state are performed by a collection of
composable model components, that set and use values in the state. For
example there is a component that evolves an equation for fluid
number density, another component that evolves pressure. These
components are configured when they are created, so that the same code
is used to evolve every species that needs that component. Every
component can access the whole state, so some perform calculations for
a single species, while others perform calculations involving multiple
species (e.g. collisions, sheath boundary conditions).

An important distinction between this design and one with a shared
global state is that here the state is an object which is passed to
components in a user-defined order. This has two advantages: It
controls when state can be modified, making the flow of the program
easier to understand, and it facilitates unit testing because the
inputs to the components can be precisely specified with no hidden
side-channels or large setup/teardown procedures.

Controlling when and how data can be modified is crucial to preventing
errors, such as components being run out of order so that quantities
are set or modified after use. The values being calculated are
physical quantities at the current simulation time, and so logically
do not change. It is therefore tempting to adopt an immutable
(persistent) data structure, such as the HAMT used in
Clojure~\cite{hickey2020} and available in C++ libraries such as
Immer~\cite{puente2017}. There are however situations in which
components need to modify fields set by earlier components; an
important example is applying boundary conditions. Rather than copy
arrays of data to maintain immutability, the approach used here is to
mark quantities as immutable after they have been used: A quantity can
be modified (e.g. boundary conditions applied) only if that quantity
has not already been used in a previous calculation.

\begin{figure}[h]
    \centering
    \includegraphics[width=0.9\textwidth]{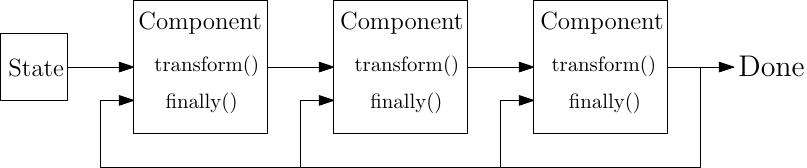}
    \caption{The State of the simulation system is passed through a
      sequence of Components in two passes. In the first pass
      components can modify the state passed to their
      \texttt{transform} function; in the second pass the state cannot
      be modified, but is used to update component internal states in
      their \texttt{finally} functions.}
    \label{fig:state-command}
\end{figure}

The flow of information carried by the state through a sequence of components
is shown in figure~\ref{fig:state-command}.
This design restricts when the state can be modified, limiting the
complexity of interactions between components, and making the logic of
the program easier to follow. In Hermes-3 the state is passed to each
component twice: The first time the state is mutable, and the
component can insert values into it. After all components have been
called in this way, the state is ``frozen'', and passed to each
component again but cannot be modified. In this second pass each
component can use the final state to update its own internal state,
such as time derivatives to be passed to the time integrator.  This
means that components can depend on each other's outputs, including
mutual dependencies, but all modifications must be in the first
function (called \texttt{transform()}), and in the second function
(called \texttt{finally()}) all components can assume that the state
will not subsequently change. This structure could be exploited to
enable all component \texttt{finally()} functions to be run in
parallel, but this is not currently done in Hermes-3.

The design of Hermes-3 enables run-time configuration of the
simulation equations: All of the examples shown in
section~\ref{sec:applications} use the same executable, despite
solving different sets of equations in different numbers of spatial
dimensions.  The overhead of this flexibility appears to be small, but
for high performance scientific simulation codes it is debatable
whether run-time configuration is essential: Once the simulation is
set up, it is performing the same set of operations repeatedly,
calculating time derivatives given different system states. Run-time
configuration enables the user (scientist) to modify the equations
without recompiling, but means that some errors are only caught at
runtime, which might have been caught more quickly at compile time.
Compile-time configuration of the equations solved might enable more
optimisations, since conditionals can be known and optimised out by
the compiler. Just-In-Time (JIT) compilation might offer the best of
both worlds; since the operations are the same with different data,
run-time analysis of the performance might enable on-the-fly tuning to
identify bottlenecks and optimise throughput.

\section{Applications}
\label{sec:applications}

We now describe some applications of Hermes-3, starting from a 1D
transport model (section~\ref{sec:1d-transport}); then a 2D transport
model in 2D (axisymmetric) tokamak geometry
(section~\ref{sec:applications-2d}).  Finally we demonstrate
time-dependent capabilities by simulating 2D (drift plane) plasma
``blobs'' (section~\ref{sec:applications-blobs}). Application of these
capabilities to 3D turbulence simulations is relatively
straightforward, but requires significantly more space to describe
adequately and so will be explored in separate publications.

\subsection{1D transport}
\label{sec:1d-transport}

We first apply Hermes-3 to a one-dimensional problem: the flow of heat
and particles along a magnetic flux tube which is in contact with a
material surface. The model includes electrons, deuterium ions and
neutral deuterium atoms. One end of the domain is modelled as being in
contact with a material surface, forming a plasma sheath and
accelerating ions to the sound speed. The flow of ions to the surface
is ``recycled'' back into the domain as neutral atoms, which then
undergo charge exchange and ionisation reactions with the plasma. The
other end of the domain has a symmetry (no-flow) boundary condition,
where thermodynamic variables (e.g. densities, pressures) have zero
gradient, and flow velocities are zero. This is a widely used model
for the divertor region of tokamak plasmas, which has several
implementations of varying
complexity~\cite{nakazawa-2000,goswami-2001,nakamura2011,togo2013,havlickova,Derks_2022},
including the SD1D model~\cite{dudson2019,bout:sd1d} which like
Hermes-3 is built on BOUT++~\cite{dudson2015}.

Despite its relative simplicity this model contains many of the
nonlinearities and numerically stiff behaviour which make 2D and 3D
plasma simulations challenging, including strong nonlinear heat
diffusion, and fast reaction rates which are sensitive to electron
temperature.

The components to be included in the simulation (section~\ref{sec:software-arch}) are specified
in an input text file; the relevant line is shown in listing~\ref{lst:hydrogen-hermes}.
\begin{lstlisting}[language=Ini,
    caption={Top-level components for 1D hydrogen transport model. Parentheses are
      used to group multi-line settings.
      Full input in \texttt{examples/1D-recycling} of the Hermes-3 repository~\cite{dudson:hermes3}},
    label={lst:hydrogen-hermes}]
[hermes]
components = (d+, d, e,
              sheath_boundary, collisions, recycling, reactions,
              electron_force_balance, neutral_parallel_diffusion)
\end{lstlisting}
The boundary condition at the material surface, implemented by the
\texttt{sheath\_boundary} component in
listing~\ref{lst:hydrogen-hermes}, is the multi-ion sheath boundary
described in~\cite{tskhakaya2005}. For the single ion species here
this reduces to the standard Bohm-Chodura-Riemann sheath boundary
condition~\cite{stangeby-1995}.  The \texttt{collisions} component
implements collisions between an arbitrary number of charged and
neutral species.  Reactions between species are organised into a
subsection called \texttt{reactions}, and are chosen to have names
which are readable and follow a convention for the species labels.
\begin{lstlisting}[language=Ini,
    caption={Reactions contained in the 1D hydrogen transport model},
    label={lst:hydrogen-hermes-reactions}]
[reactions]
type = (
        d + e -> d+ + 2e,     # Deuterium ionisation
        d+ + e -> d,          # Deuterium recombination
        d + d+ -> d+ + d,     # Charge exchange
        )
\end{lstlisting}
Reaction cross-sections for hydrogen and helium have been taken from
the Amjuel database~\cite{amjuel}. 

Each particle species has components to evolve the density, pressure
and parallel momentum, and a no-flow boundary condition imposed on the
upstream boundary.  To illustrate how Hermes-3 components combine to
form the equations solved, the \texttt{d+} (deuterium ion) species
settings are shown in listing~\ref{lst:hydrogen-hermes-d+}.
\begin{lstlisting}[language=Ini,
    caption={Components to model the deuterium ion species},
    label={lst:hydrogen-hermes-d+}]
[d+]  # Deuterium ions
type = (evolve_density, evolve_pressure, evolve_momentum,
        noflow_boundary, upstream_density_feedback)
charge = 1  # charge
AA = 2      # mass [amu]
density_upstream = 1e19 # Upstream density [m^-3]
recycle_as = d          # Species to recycle as
recycle_multiplier = 1  # Recycling fraction
\end{lstlisting}
This implements a set of equations for the density $n_{d+}$, pressure
$p_{d+} = en_{d+}T_{d+}$ and parallel velocity $v_{||d+}$ of the
deuterium ions (\texttt{d+}) of mass $m_{d+}$ and charge $q_{d+}$,
given in equation~\ref{eq:1d-hydrogen-d+}. These equations are in SI
units except temperatures in eV.
\begin{subequations}
  \label{eq:1d-hydrogen-d+}
  \begin{align}
    \frac{\partial n_{d+}}{\partial t} &= -\nabla\cdot\left(n_{d+}\mathbf{b}v_{||d+}\right) \nonumber \\
    & + \underbrace{S_{\textrm{PI}}}_{\texttt{upstream\_density\_feedback}} + \underbrace{n_en_d\left<\sigma v\right>_{iz}}_{\texttt{d + e -> d+ + 2e}} - \underbrace{n_en_{d+}\left<\sigma v\right>_{rc}}_{\texttt{d+ + e -> d}} \\
    \frac{\partial p_{d+}}{\partial t} &= -\nabla\cdot\left(p_{d+}\mathbf{b}v_{||d+}\right) - \frac{2}{3}p_{d+}\nabla\cdot\left(\mathbf{b}v_{||d+}\right) + \nabla\cdot\left(\kappa_{||d+}\mathbf{b}\mathbf{b}\cdot\nabla T_{d+}\right) \nonumber \\
    & + \underbrace{n_en_d\left<\sigma v\right>_{iz}\left[eT_d + \frac{1}{2}m_d\left(v_{||,d} - v_{||,d+}\right)^2\right]}_{\texttt{d + e -> d+ + 2e}} \nonumber \\
    & + \underbrace{n_{d+}n_{d}\left<\sigma v\right>_{cx}\left[e\left(T_d - T_{d+}\right) + \frac{1}{2}m_d\left(v_{||,d} - v_{||,d+}\right)^2\right]}_{\texttt{d + d+ -> d+ + d}} \nonumber \\
    & - \underbrace{n_en_{d+}\left<\sigma v\right>_{rc}eT_{d+}}_{\texttt{d+ + e -> d}} + \underbrace{W_{d+}}_{\texttt{collisions}} \\
    \frac{\partial}{\partial t}\left(m_{d+}n_{d+}v_{||d+}\right) &= -\nabla\cdot\left(m_{d+}n_{d+}v_{||d+}\mathbf{b}v_{||d+}\right) - \mathbf{b}\cdot\nabla p_{d+} + q_{d+}n_{d+}E_{||}\nonumber \\
    & + \underbrace{n_en_d\left<\sigma v\right>_{iz}m_dv_{||d}}_{\texttt{d + e -> d+ + 2e}} - \underbrace{n_en_{d+}\left<\sigma v\right>_{rc}m_{d+}v_{||d+}}_{\texttt{d+ + e -> d}} \\
    & + \underbrace{n_{d+}n_{d}\left<\sigma v\right>_{cx}m_{d+}\left(v_{||d} - v_{||d+}\right)}_{\texttt{d + d+ -> d+ + d}} + \underbrace{F_{d+}}_{\texttt{collisions}}
  \end{align}
\end{subequations}
Where $\mathbf{b}\equiv\mathbf{B}/B$ is the unit vector in the
direction of the magnetic field $\mathbf{B}$, and $\kappa_{||d+}$ is
the parallel heat conduction coefficient that depends on the collision
frequency calculated by the \texttt{collisions} component.  For each
equation~\ref{eq:1d-hydrogen-d+} the first line corresponds to the
essential transport terms implemented in the \texttt{evolve\_density},
\texttt{evolve\_momentum} and \texttt{evolve\_pressure} components.
Additional components, labelled with underbraces in
equation~\ref{eq:1d-hydrogen-d+}, add sources and sinks that modify
and couple species together.  Details of the full system of 7 evolving
equations are given in~\ref{sec:equations}.

\begin{figure}[h]
  \centering
  \includegraphics[width=0.9\textwidth]{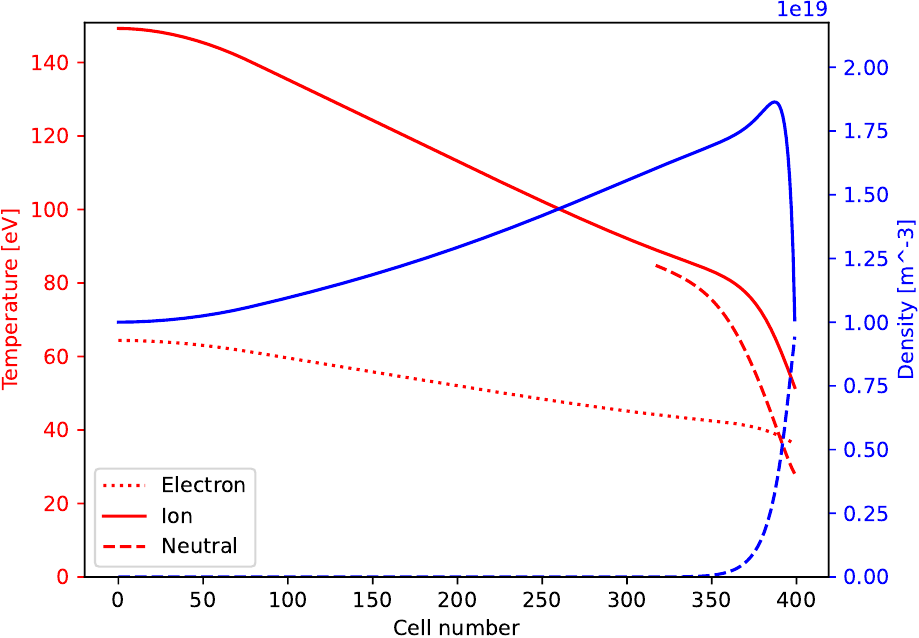}
  \caption{Steady-state solution to system of
    equations~\ref{eq:1d-hydrogen-d+} and \ref{sec:equations} in one
    dimension. 50MW of power enters a source region on the left,
    driving plasma-neutral interactions including ionisation, leaving
    through the sheath boundary on the right. 100\% of the plasma ions
    leaving the right boundary are recycled as neutral
    atoms. Simulation inputs are in \texttt{examples/1D-recycling} in
    the Hermes-3 repository.}
  \label{fig:hydrogen_solution}
\end{figure}

\begin{figure}[h]
  \centering
  \includegraphics[width=0.9\textwidth]{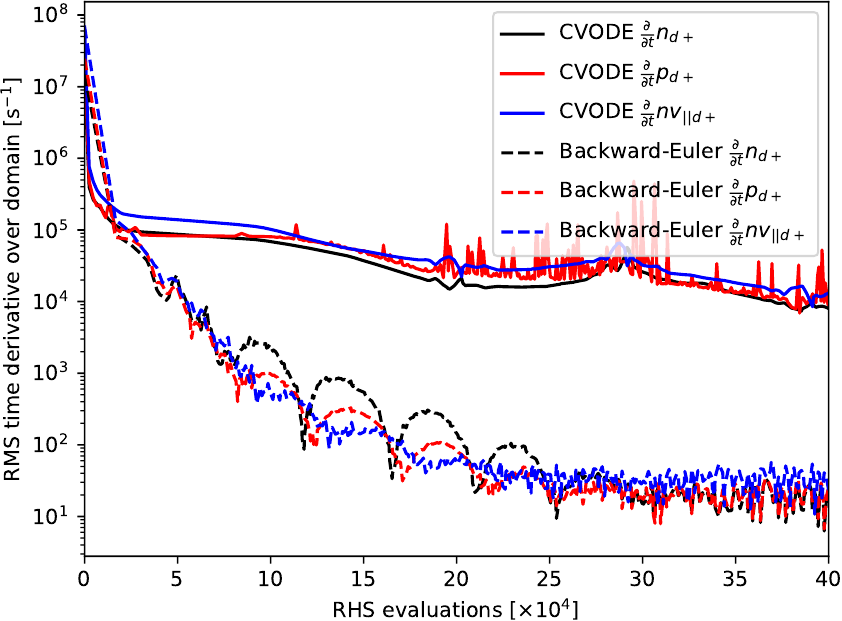}
  \caption{Root-mean-square time derivatives of deuterium density
    (Nd+), pressure (Pd+) and momentum (NVd+) as a function of
    iteration (RHS evaluation). These converge towards zero as the
    system approaches steady state. Results are shown for Backward
    Euler Newton-Krylov with Jacobian coloring and iLU preconditioning
    (NK); and the CVODE time integrator. Figure produced by
    \texttt{examples/1D-recycling/plot\_convergence.py}.}
  \label{fig:timederivs-rhsevals}
\end{figure}

The equations are integrated in time towards a steady state solution
using the backward Euler method described in
section~\ref{sec:time-integration}. The result is shown in
figure~\ref{fig:hydrogen_solution}.
The root-mean-square of the time-derivatives of ion density, pressure
and parallel momentum are shown in
figure~\ref{fig:timederivs-rhsevals} as a function of the number of
right-hand-side (RHS) evaluations, a measure of the computational
cost. This evaluation count includes those performed as part of the
finite difference Jacobian approximation. This shows a reduction in
the time derivatives of the system by almost six orders of magnitude
in $10^5$ RHS evaluations. For the above system of equations, with 400
grid cells, this calculation takes approximately 30 minutes on a single
core.
For comparison the convergence towards steady state with the Sundials
CVODE~\cite{hindmarsh2005} library is shown in
figure~\ref{fig:timederivs-rhsevals}.  CVODE uses an adaptive order,
adaptive timestep Backward Differentiation Formula (BDF) method, and
is highly effective for time-dependent problems of interest even
without preconditioning (e.g.  the plasma blobs application,
section~\ref{sec:applications-blobs}). For this problem only the
parallel heat conduction is preconditioned.
Figure~\ref{fig:timederivs-rhsevals} shows that the Backwards-Euler
with Jacobian coloring preconditioner method can provide significantly
better performance for steady-state problems, though it would not be a
good choice for time-dependent simulations, being only first order
accurate in time.

\begin{figure}[h]
  \centering
  \includegraphics[width=0.9\textwidth]{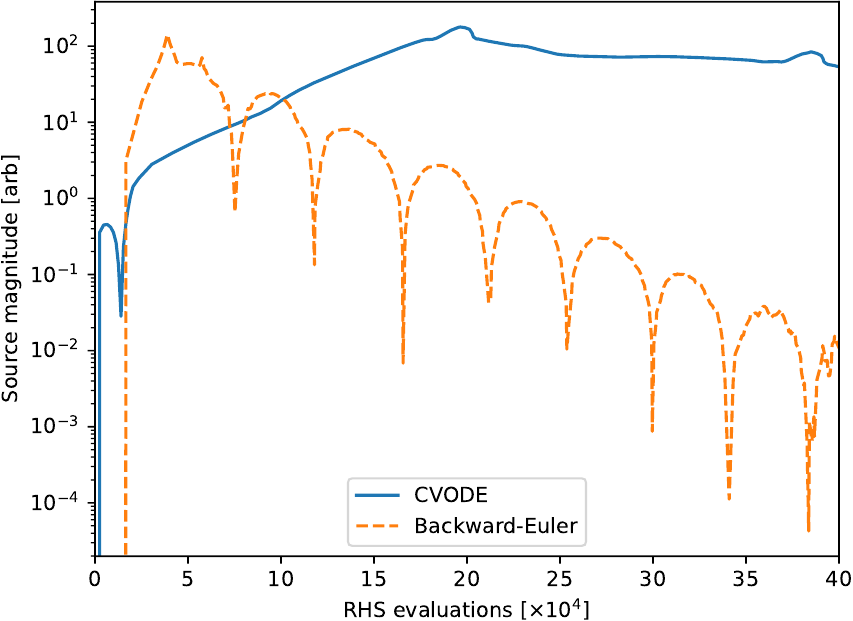}
  \caption{Convergence of the density source with RHS evaluation in a
    1D simulation with 100\% recycling
    (figure~\ref{fig:hydrogen_solution}) where the true steady state
    source is zero. Figure produced by
    \texttt{examples/1D-recycling/plot\_convergence.py}.}
  \label{fig:source_rhsevals}
\end{figure}

In this simulation the recycling at the ``target'' end of the domain
was set to 100\%, while there is a no-flow condition on the upstream
boundary. A Proportional-Integral (PI) controller is used to control
an upstream particle source (implemented in the
\texttt{upstream\_density\_feedback} component); As the target
upstream density is approached, this input source should go to zero if
mass flow is conserved.  Figure~\ref{fig:source_rhsevals} shows that
this does indeed happen: The source converges towards exponentially
towards zero in steady state.

The 1D simulation described here is a useful tool in its own right, for
studies of plasma dynamics and detachment in magnetised plasmas. The
advantage of the software design used in Hermes-3
(section~\ref{sec:software}) is that the same code can extend to more
complex models and higher dimensions with only changes to the input.

\subsubsection{Impurity seeding}

We now extend the 1D simulation described in
section~\ref{sec:1d-transport} to multiple ion species, by including
all ten charge states of neon as separate species.  The simulation now
contains 40 evolving fields: The density, pressure and momentum of all
deuterium and neon ion and atomic charge states (13 ion species in
total), and the electron pressure. These species are coupled through
collisions, thermal forces, the parallel electric field, and 32 atomic
reactions: ionisation, 3-body recombination and charge exchange
recombination (with deuterium ions) of each ionisation level of neon;
ionisation and charge exchange of neutral deuterium atoms to deuterium
ions. The modular structure of the code
(section~\ref{sec:software-arch}) enables this to be accomplished
relatively straightforwardly by changing the input file.

\begin{lstlisting}[language=Ini,
    caption={Top-level components for 1D transport model with neon.
      Input \texttt{examples/1D-neon}},
    label={lst:neon-hermes}]
[hermes]
components = (d+, d, ne, ne+, ne+2, ne+3, ne+4, ne+5, ne+6,
              ne+7, ne+8, ne+9, ne+10, e, sheath_boundary,
              thermal_force, collisions, recycling, reactions,
              electron_force_balance, neutral_parallel_diffusion)
\end{lstlisting}
There are now many more reactions, but the input remains clear:

\begin{lstlisting}[language=Ini,
    caption={Reactions contained in the 1D transport model with neon.},
    label={lst:neon-hermes-reactions}]
[reactions]
type = (
        d + e -> d+ + 2e,     # Deuterium ionisation
        d + d+ -> d+ + d,     # Charge exchange

        ne + e -> ne+ + 2e,   # Neon ionisation
        ne+ + e -> ne,        # Neon+ recombination
        ne+ + d -> ne + d+,   # Neon+ charge exchange recombination

        ...
        )
\end{lstlisting}
The cross-sections and radiated power from the neon reactions are
calculated using ADAS~\cite{adas}: \texttt{scd96} and \texttt{plt96} for
ionisation; \texttt{acd96} and \texttt{prb96} for recombination;
\texttt{ccd89} for charge exchange. These files were converted
to JSON format using atomic++~\cite{atomicpp}.

Collisions and the thermal forces between species are calculated as
described in section~\ref{sec:1d-transport}. Those Braginskii energy
and momentum exchange rates are approximations which are only strictly
valid when heavy ions are trace impurities. In the simulations shown
here the neon concentration is small (a fraction of a percent). More
complete models of collisions in a multi-ion plasma have been
derived~\cite{zhdanov2002} and recently
generalised~\cite{raghunathan2021}. Implementing these models into
Hermes-3 is left as future work, but is not anticipated to present any
fundamental difficulty.

\begin{figure}[h]
  \centering
  \includegraphics[width=0.9\textwidth]{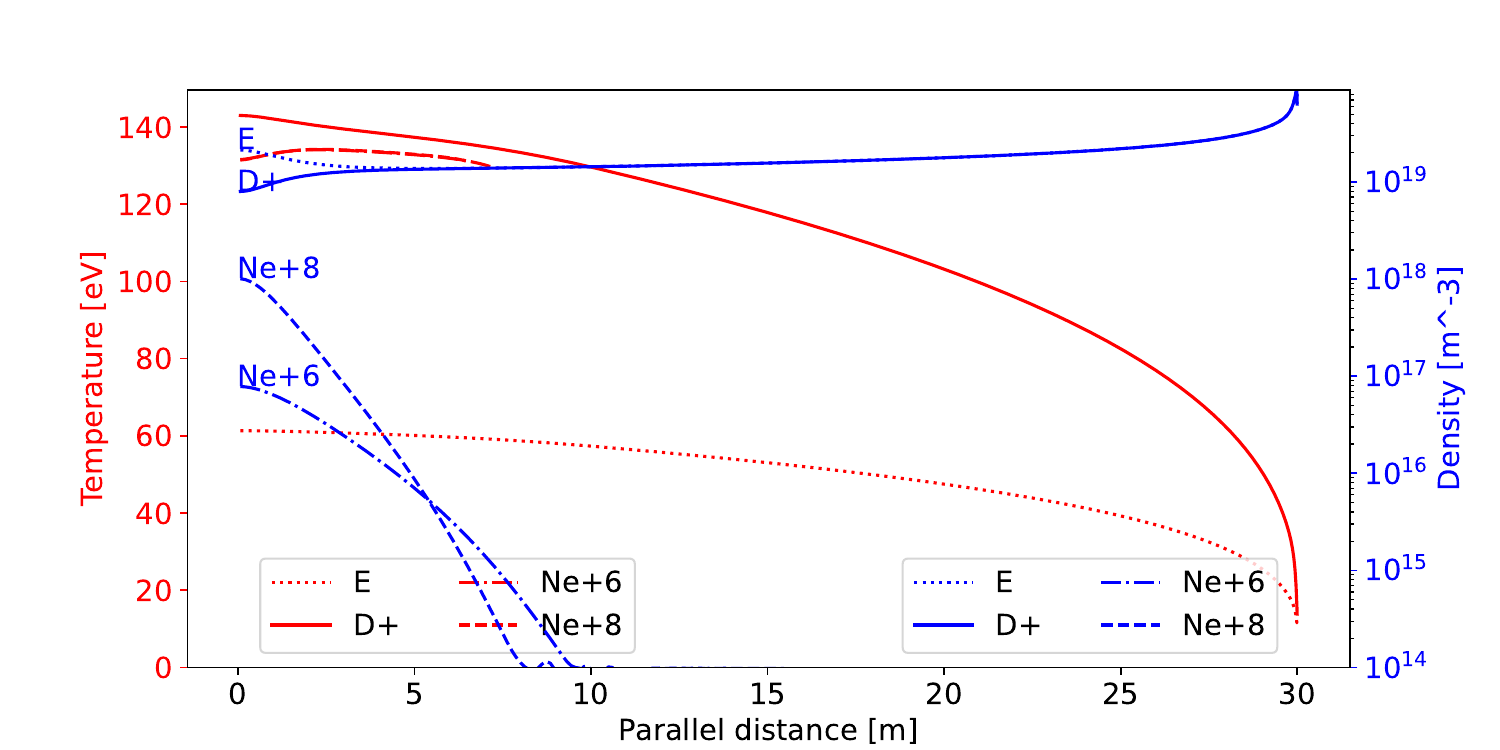}
  \caption{Steady state solution with 100\% recycling, evolving all
    charge states of neon as separate fluids with their own densities,
    temperatures and flow velocities. A subset of species densities
    (blue lines) are shown on a logarithmic scale. Simulation inputs
    in \texttt{examples/1D-neon} of the Hermes-3 repository.}
  \label{fig:1d-neon-steadystate}
\end{figure}

Starting from the hydrogen simulation described above
(section~\ref{sec:1d-transport}), a simulation with 100\% recycling
and an initial uniform low concentration of neon is run to steady
state, and shown in figure~\ref{fig:1d-neon-steadystate}.  In this
simulation there is no net flow upstream of the ionisation region, and
so thermal forces drive neon impurities upstream.
The steady-state solution is shown in
figure~\ref{fig:1d-neon-steadystate}. Future applications of this
capabilty include simulating impurity-seeded plasma detachment
phenomena.

\subsection{2D (axisymmetric) transport}
\label{sec:applications-2d}

The same code that is used in a 1D domain in the previous sections can
be applied to 2D tokamak domains with one or two X-points. By
introducing cross-field diffusion of both charged and neutral species,
an axisymmetric tokamak transport simulation in the spirit of
SOLPS~\cite{schneider-2006}, EDGE2D~\cite{simonini-1994} or
UEDGE~\cite{rognlien-2002} can be performed, though not yet at a
comparable level of maturity or completeness.  To demonstrate the
ability of Hermes-3 to solve axisymmetric transport problems,
simulations are performed with deuterium ions and neutral atoms.
Diffusion coefficients and plasma parameters are taken
from~\cite{hromasova-2021,hromasova-thesis}: Spatially constant
cross-field diffusion coefficients for particle transport
$D_n=0.15$m$^2/$s, electron and ion thermal transport $\chi_e = \chi_i
= 4$m$^2/$s. In general these coefficients can be functions of
location, and can be different for each species.

\begin{figure}
  \centering
  \begin{subfigure}[h]{0.48\textwidth}
    \centering
    \includegraphics[width=\textwidth]{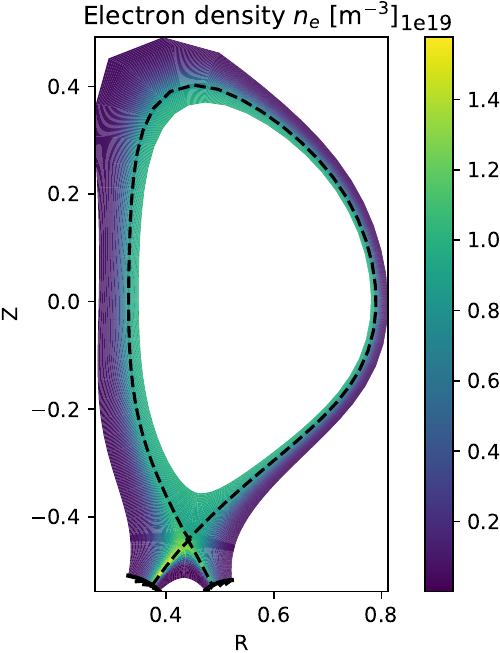}
    \caption{Electron density. The core boundary is fixed to a density
      of $1\times 10^{19}$m$^{-3}$. At divertor targets 99\% of ion
      flux is recycled as neutral atoms.}
    \label{fig:2d_ne}
  \end{subfigure}
  \hfill
  \begin{subfigure}[h]{0.48\textwidth}
    \centering
    \includegraphics[width=\textwidth]{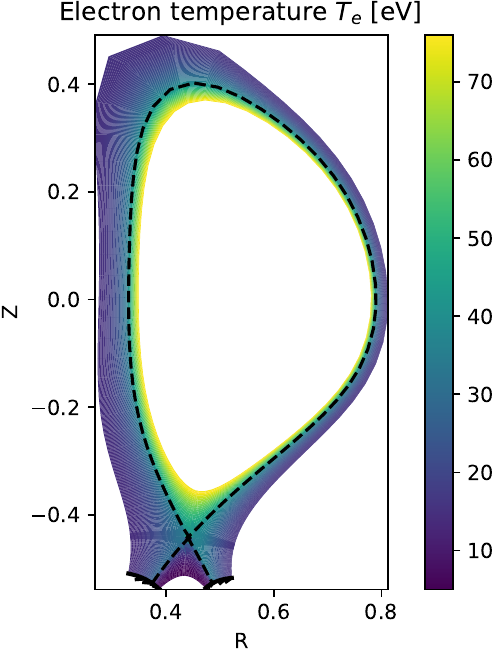}
    \caption{Electron temperature. At the core boundary both electron and ion temperatures
      are fixed to $75$eV.}
    \label{fig:2d_te}
  \end{subfigure}
  \caption{Axisymmetric tokamak transport simulation of deuterium ions
    and atoms. Equations given in~\ref{sec:equations}. Simulation inputs
    in \texttt{examples/tokamak/recycling} of the Hermes-3 repository.}
  \label{fig:2d_domain}
\end{figure}

The plasma equilibrium is based on a COMPASS-like equilibrium
generated using analytic Grad-Shafranov solutions~\cite{cerfon:2010,
  omotani:cfg}.  The domain simulated is shown in
figure~\ref{fig:2d_domain}, consisting of a narrow annulus around the
separatrix (dashed black lines in figure~\ref{fig:2d_domain})
including closed and open field line regions. The radial boundaries
are at normalised poloidal flux ($\psi_N$) of 0.9 in the core and in the private flux
region (PFR), and 1.3 in the Scrape-Off Layer (SOL). The Hypnotoad
tool~\cite{hypnotoad} was used to generate a sequence of grids of
increasing resolution from $16\times 24$ to $64\times 96$ (radial
$\times$ poloidal cells).

As in previous examples, the equations solved are specified as a set
of components:
\begin{lstlisting}[language=Ini,
    caption={Top-level components for 2D transport model.
      Full input in \texttt{examples/tokamak/recycling} of the Hermes-3 repository.},
    label={lst:hermes-2d-components}]
components = (d+, d, e,
              collisions, sheath_boundary_simple, 
              recycling, sound_speed, reactions,
              electron_force_balance)
\end{lstlisting}
The deuterium ion species is configured with a set of components
representing the equations solved, given in
listing~\ref{lst:hermes-2d-d1}

\begin{lstlisting}[language=Ini,
    caption={Deuterium ion components for 2D transport model},
    label={lst:hermes-2d-d1}]
[d+]
type = (evolve_density, evolve_momentum, evolve_pressure,
        anomalous_diffusion)
anomalous_D = 0.15   # Density diffusion [m^2/s]
anomalous_chi = 4    # Thermal diffusion [m^2/s]
...
\end{lstlisting}
which is similar to the configuration in 1D simulations given in
listing~\ref{lst:hydrogen-hermes-d+}, but adds anomalous cross-field
diffusion terms.  Reactions between species are calculated using
Amjuel rates~\cite{amjuel}, comprising ionisation, recombination, and
charge exchange processes as described in
section~\ref{sec:1d-transport}.

At the inner (core) boundary the deuterium density is fixed to
$1\times 10^{19}$m$^{-3}$; electron and ion temperatures are set to
$75$eV.  This core boundary therefore acts as a source of heat and
particles. At the target plates a sheath boundary condition is applied
in which the plasma flow goes to the sound speed, with a recycling
fraction of 0.99 so that there is a flux of neutral atoms into the
domain at the target plates. The 1\% of ion flux that is not recycled
is balanced in steady state by a diffusion of ions from the core
boundary. This particle flux balance will be used in
section~\ref{sec:2d_convergence} to verify the conservation of
particles in these simulations.

The heat flux along the magnetic field into the target plates is given
by $q_{||e,i} = \gamma_{e,i}neTc_s$ with sheath heat transmission
factors $\gamma_e=4.8$ for electrons and $\gamma_i=3.5$ for ions. The
sound speed into the sheath is $c_s = \sqrt{e\left(T_i +
  T_e\right)/m_i}$.  There are no diffusive fluxes to the outer walls
because zero-gradient boundary conditions are used there. The power
into the target plates should therefore equal the input power through
the core boundary, less the power radiated during atomic processes
(primarily ionisation). This is used in
section~\ref{sec:2d_convergence} to assess conservation of energy.

The full set of equations solved are given in \ref{sec:equations}.

\subsubsection{Evolution to steady state}

\begin{figure}
  \centering
  \begin{subfigure}[h]{0.49\textwidth}
    \centering
    \includegraphics[width=\textwidth]{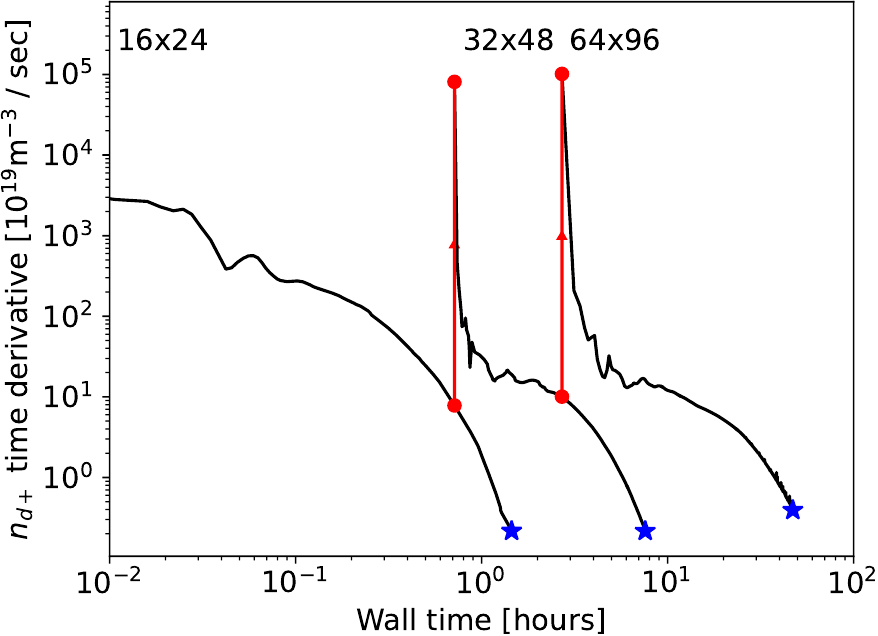}
    \caption{CVODE solver with heat conduction preconditioner}
    \label{fig:2d_history_cvode}
  \end{subfigure}
  \hfill
  \begin{subfigure}[h]{0.49\textwidth}
    \centering
    \includegraphics[width=\textwidth]{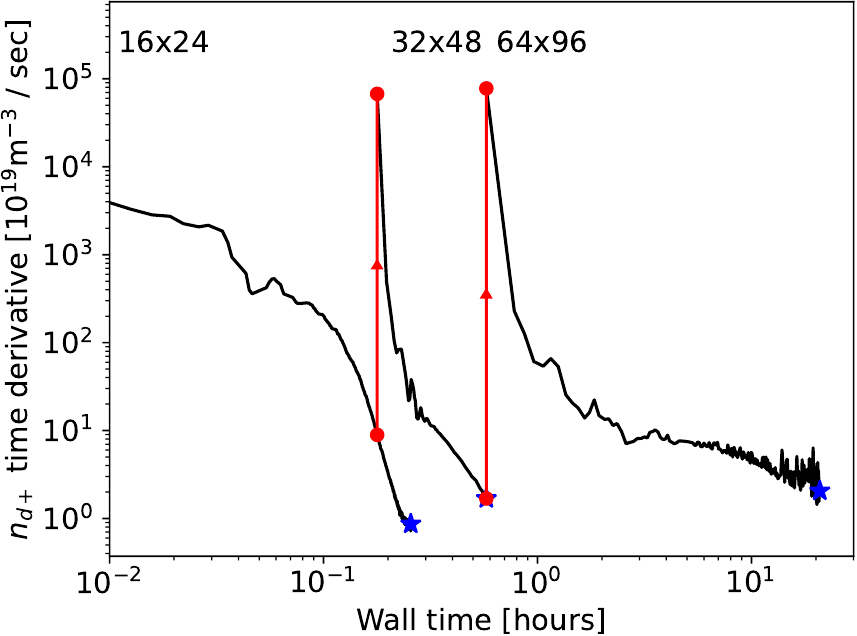}
    \caption{Backward Euler solver with PETSc matrix coloring and Hypre's Euler preconditioner}
    \label{fig:2d_history_beuler}
  \end{subfigure}
  \caption{Evolution of the Root-Mean-Square (RMS) time derivative
    residuals. Vertical red arrows indicate where the solution is
    interpolated onto a higher resolution mesh. Blue stars are the
    solutions that are compared in section~\ref{sec:2d_convergence}.
    The number of cores used is increased with the grid resolution:
    $12$ ($16\times 24$ mesh), $48$ ($32\times 48$ mesh) and $192$
    ($64\times 96$ mesh).}
  \label{fig:2d_history}
\end{figure}

The system of transport equations is relatively small (e.g. 10,752
variables for the $32\times 48$ mesh) but highly nonlinear and with a
wide range of timescales, making finding steady state solutions
challenging. Simple application of a nonlinear solver does not
converge in most cases of interest, and the system must be regularised
using a (pseudo-)timestepping approach. As the system approaches
steady state the timestep can be made progressively larger, usually in
an automated manner based on the number of nonlinear iterations
required to converge the previous step. The combination of time
integration method, time step adjustment heuristic, nonlinear solver,
inner linear iterative solver, and preconditioner have many parameters
that can affect performance. The nested methods interact in ways that
are problem-dependent, making general conclusions regarding
performance difficult to draw. The performance results presented here
(e.g. figure~\ref{fig:2d_history}) should therefore be taken as
somewhat anecdotal, the important aspects being the energy and
particle conservation properties (table~\ref{tab:balances}) and
scaling with problem size.

In general power balance reaches steady-state on a shorter timescale
than the particle balance: The thermal energy content of the system
(plasma + neutrals) is $W\simeq 47$J, so the energy confinement time
is $\tau_E \equiv W / P_{in}\simeq 0.24$ms. On the other hand the ion
particle content is approximately $2\times 10^{18}$, giving a particle
throughput timescale of $\tau_p\simeq 45$ms. This longer particle
balance timescale becomes increasingly challenging at high recycling
fractions relevant to large fusion devices.

An effective strategy, already used routinely and interactively in
UEDGE, is progressive mesh refinement: Starting on the coarsest mesh
($16\times 24$ here), CVODE is used with an absolute tolerance of
$10^{-12}$ and relative tolerance $10^{-5}$, tightening the relative
tolerance to $10^{-8}$ as steady state is approached. These tolerances
can be loosened in some cases, but at the risk of numerical
instability and convergence failure after a number of steps.  Once
progress has been made on a coarse mesh, the solution is interpolated
onto a higher resolution mesh (using SciPy's
RegularGridInterpolator~\cite{2020SciPy-NMeth} over logically
rectangular mesh patches). The simulation is then continued using an
increased number of cores. The refinement process may be repeated.
Figure~\ref{fig:2d_history} shows the Root-Mean-Square (RMS) of the
time derivatives of the plasma density averaged over the domain, as a
function of wall clock time (running on NERSC's Perlmutter).
For each mesh resolution the simulation was continued after
interpolation, until the Root-Mean-Square time scale exceeded one
second, to minimise the impact of mesh interpolation error on the
comparison of solutions. As the mesh was refined the number of cores
used was increased following a weak scaling. The increase in run time
with grid resolution in figure~\ref{fig:2d_history} is primarily
driven by the number of iterations required: For CVODE the iteration
counts are $3.1\times 10^6$ ($16\times 24$ mesh), $1.2\times 10^7$
($32\times 48$ mesh) and $3.7\times 10^7$ ($64\times 96$ mesh). The
time per iteration (RHS evaluation) is 1.7ms, 2.0ms and 4.3ms
respectively.

Figure~\ref{fig:2d_history_beuler} shows results using the Backward
Euler method with Jacobian coloring and Euler ILU preconditioner that
was described in section~\ref{sec:time-integration} and used in 1D
simulations (section~\ref{sec:1d-transport}). In general a significant
improvement in performance is found, primarily due to a reduction in
the number of iterations required. The speedup is most significant
for small meshes and processor counts, where direct ILU preconditioners
would be expected to have an advantage over matrix-free methods, but
that advantage to reduces for large meshes and processor counts.

\subsubsection{Convergence and accuracy}
\label{sec:2d_convergence}

\begin{figure}
  \centering
  \includegraphics[width=\textwidth]{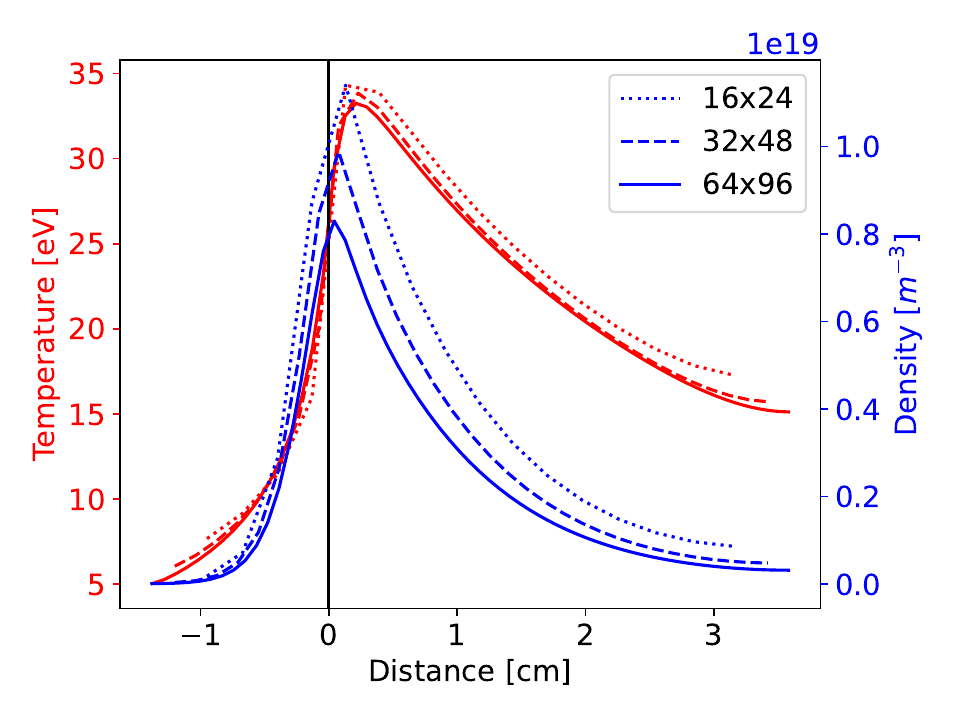}
  \caption{Electron temperature and density at the outer target.
    Dotted: $16\times 24$ resolution; Dashed: $32\times 48$; Solid: $64\times 96$.}
  \label{fig:2d_outer_target}
\end{figure}

The accuracy of the methods are now assessed by examining the
conservation properties and convergence of the solution with mesh
resolution. Figure~\ref{fig:2d_outer_target} shows the profiles of
density and temperature along the outer target, for each mesh
resolution.
Low resolution meshes broaden the profiles of both density and
temperature relative to high resolution cases: Numerical dissipation
enhances the effective cross-field diffusion. Given the fixed
(Dirichlet) core boundary conditions, this enhanced diffusion
increases the power into the domain at low resolution.  It also be
seen in figure~\ref{fig:2d_outer_target} that the meshes do not have a
consistent boundary location: Due to boundary cell locations, as the
grid is refined the outer edges of the domain converge at first order
to the specified poloidal flux values.  This will limit the global formal
convergence to at best first order unless improvements are made to the
mesh generator.

To assess power and particle balance, table~\ref{tab:balances} lists
the flows of power and particles into and out of the domain, for each
mesh resolution.
\begin{table}[h!]
  \centering
  \caption{Global power and particle balance in 2D transport simulations}
  \label{tab:balances}
  \begin{tabular}{ l c c c }
    & $16\times 24$ & $32\times 48$ & $64\times 96$ \\ 
    \hline
    Input power [kW]                   & 195.6  & 174.9  & 162.9 \\
    Power to outer target [kW]         & 104.6  &  92.5  &  85.9 \\
    Power to inner target [kW]         &  75.4  &  64.9  &  59.5 \\
    Power to atomics [kW]              &  22.8  &  19.3  &  17.6 \\
    Power balance error [kW]           &   7.2 (3.7\%)  &   1.7  (0.98\%) &   0.092 (0.056\%) \\
    \hline
    Input ion flux [$10^{19}$/s]        &  5.31  &  4.56  & 4.15 \\
    Flux to outer target [$10^{19}$/s]  &   278  &   247  &  230 \\
    Flux to inner target [$10^{19}$/s]  &   284  &   229  &  204 \\
    Recycling fraction [0.99]          & 0.9909 & 0.9903 & 0.9904
  \end{tabular}
\end{table}
Power enters the domain through the inner (core) boundary, where
Dirichlet boundary conditions are set on density and temperatures so
that power crossing this boundary depends on the local gradients. The
target temperatures in figure~\ref{fig:2d_outer_target} are well above
the $5$eV typical for detachment, so these simulations are in attached
conditions and most of the power goes to the outer and inner targets.
Some power is lost through atomic processes, both to overcome
ionisation potentials and through radiation. Power to atomics includes
the deuterium ionisation potential so this potential energy flux is
not included in the power to outer and inner targets listed in
table~\ref{tab:balances}.  As noted in
section~\ref{sec:finite_differencing} pressure equations are evolved
rather than energy, so that energy conservation is in general not
exact but converges as the mesh is refined. For comparison, a 1\%
power balance error has been used as a SOLPS-ITER convergence
criterion~\cite{WIESEN2015480}.

Particle fluxes are shown in Table~\ref{tab:balances} as the flux into
the domain through the inner boundary, and the fluxes to inner and
outer targets. Due to the imposed recycling fraction of 0.99, we
expect 1\% of the flux to the targets to be lost (pumped), and
replaced by a matching flux of ions into the domain from the core. The
core and target fluxes are therefore used to infer the recycling
fraction in Table~\ref{tab:balances}. If particle balance is achieved
then that fraction should match the 0.99 value set. We find this to be
well matched: Particle conservation is significantly easier to achieve
in this system of equations than energy conservation, and these
results demonstrate that all advection and diffusion operators,
recycling and atomic processes, properly conserve particle fluxes.

\subsection{2D (drift-plane) blobs}
\label{sec:applications-blobs}

We now turn from steady-state transport problems to time-dependent
problems involving an evolving vorticity equation and electrostatic
potential $\phi$. The development of this capability towards full 3D
turbulence, particularly in the presence of multiple ion species, will
be the subject of a future publication. As an initial step and proof
of principle, we present here some examples of 2D drift-plane
simulations of plasma ``blobs'' or filaments.

The significant lines in the input file which configure this model are
shown in listing~\ref{lst:blob2d}.
\begin{lstlisting}[language=Ini,
    caption={Component configuration for isothermal blob
      simulation. Full input in \texttt{examples/blob2d} of the Hermes-3 repository.},
    label={lst:blob2d}]
[hermes]
components = e, vorticity, sheath_closure

[e]  # Electrons
type = evolve_density, isothermal
charge = -1
AA = 1/1836      # Mass of species [amu]
temperature = 5  # Temperature in eV

[sheath_closure]
connection_length = 10 # meters
\end{lstlisting}
These set up components for the electron species density and
(isothermal) temperature, a vorticity equation, and a model
for the divergence of parallel current due to the sheath closure.
This corresponds to model equations
\begin{subequations}
  \label{eq:blob2d}
\begin{align}
  \frac{\partial n_e}{\partial t} &= - \nabla\cdot\left(n_e\frac{1}{B}\mathbf{b}\times\nabla\phi\right) + \underbrace{\nabla\cdot{\frac{1}{e}\mathbf{j}_{sh}}}_{\texttt{sheath\_closure}} \\
  p_e &= \underbrace{e n_e T_e}_{\texttt{isothermal}} \\
  \frac{\partial\Omega}{\partial t} &= - \nabla\cdot\left(\Omega \frac{1}{B}\mathbf{b}\times\nabla\phi\right) + \nabla\left(p_e\nabla\times\frac{\mathbf{b}}{B}\right) + \underbrace{\nabla\cdot\mathbf{j}_{sh}}_{\texttt{sheath\_closure}} \\
  &\nabla\cdot\left(\frac{\overline{m_in}}{B^2}\nabla_\perp\phi \right) = \Omega \label{eq:potential}
\end{align}
\end{subequations}
where $n_e$ is the electron density, $p_e$ the pressure and $T_e$ the
(fixed) temperature.  The Boussinesq approximation is used here, so
the potential $\phi$ is calculated from vorticity $\Omega$ using
equation~\ref{eq:potential} with a constant mass density
$\overline{m_in}$. The divergence of current density to the sheath is
$\nabla\cdot \mathbf{j}_{sh} = n_e\phi / L_{||}$ where $L_{||}$ is the
connection length ($10$m here).

The scaling of sheath-connected isothermal plasma blobs with blob size
is a well known test case, which can be derived analytically in the
limits of large blobs where sheath current balances the divergence of
diamagnetic current, and for small blobs where polarisation current
balances the divergence of diamagnetic current (see
e.g.~\cite{omotani-2015}). The blob size $\delta$ for which the
divergence of polarisation and sheath currents contribute
approximately equally is denoted $\delta^*$.

\begin{figure}[h]
  \centering
  \begin{subfigure}[h]{0.47\textwidth}
    \centering
    \includegraphics[width=\textwidth]{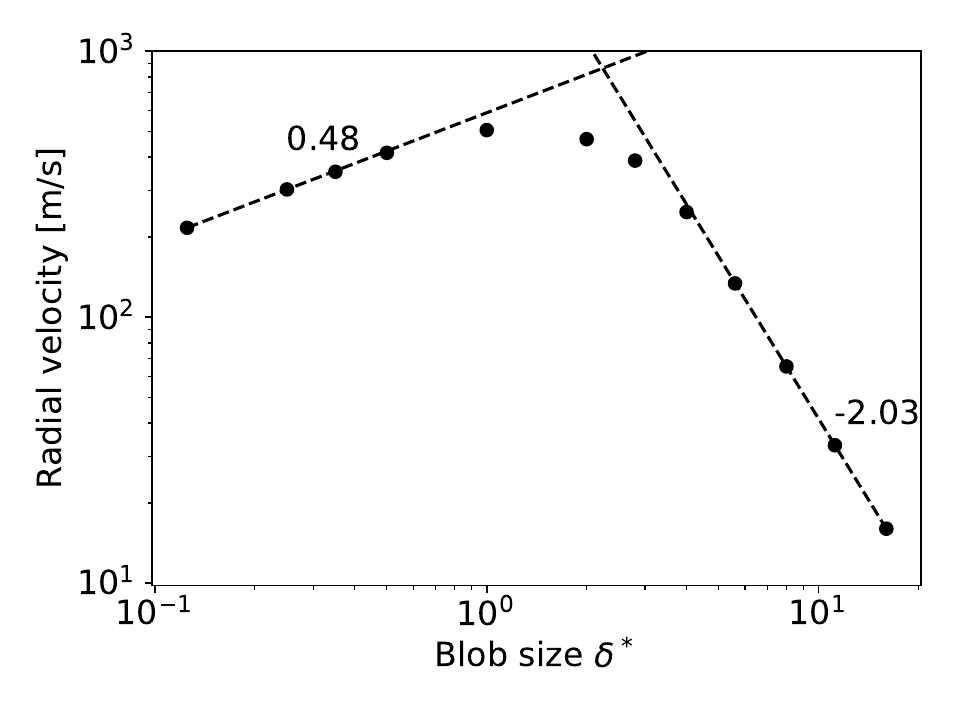}
    \caption{Radial velocity $v_r$ of an isothermal plasma blob as a
      function of blob size $\delta$. Analytical scalings are
      $v_r\sim\sqrt{\delta}$ for $\delta/\delta^* \ll 1$, and
      $v_r\sim\left(\delta / \delta^*\right)^{-2}$ for $\delta /
      \delta^* \gg 1$.}
    \label{fig:blob-velocity}
  \end{subfigure}
  \begin{subfigure}[h]{0.47\textwidth}
    \centering
    \includegraphics[width=\textwidth]{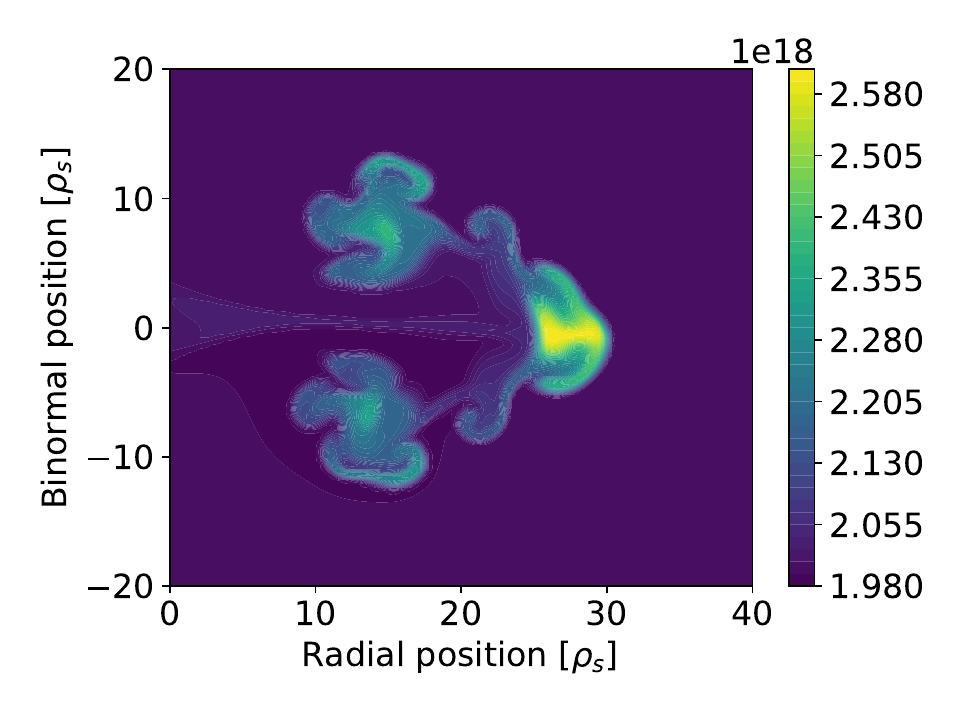}
    \caption{Electron density $n_e$ at $t=1500/\omega_{ci} = 44.7\mu$s
      for a $\delta^*$-sized seeded blob.}
    \label{fig:blob2d_density}
  \end{subfigure}
  \caption{Solution to equations~\ref{eq:blob2d} in a 2D domain
    perpendicular to the magnetic field, starting with a circular
    cross section density perturbation and driven by magnetic field
    curvature.  Input and analysis scripts in Hermes-3 repository
    \texttt{examples/blob2d}.}
  \label{fig:blob2d}
\end{figure}

Simulations are started with a circular Gaussian density perturbation
(a plasma ``blob''), whose size perpendicular to the magnetic field
and the size of the simulation domain is varied.  Because we are
interested in time-varying solutions to these equations (propagation
of plasma blobs), the CVODE time integrator~\cite{hindmarsh2005} is
used, not the backward Euler method used in
section~\ref{sec:1d-transport}.  The result is shown in
figure~\ref{fig:blob-velocity}, reproducing well known scaling of
plasma blob velocity with blob size~\cite{omotani-2015}.

This model extends quite straightforwardly to include hot ion effects
and separate ion and electron temperatures, by modifying the input
file to introduce a new species \texttt{h+} with a separate pressure
equation. The vorticity formulation is implemented such that the
polarisation current contribution of multiple species is included in
calculating the electric field; The self-consistent calculation of the
polarisation drift on the ion species density in a multi-ion species
calculation has recently been implemented and is being tested. Further
examples, tests and applications may be found in the Hermes-3
manual~\cite{dudson:hermes3-manual} and source code
repository~\cite{dudson:hermes3}.

\section{Conclusions}
\label{sec:conclusions}

Advancing understanding of the physics of the edge of tokamak plasmas
drives the need for increasingly complex models. To address this need
a new open-source plasma simulation tool has been developed that
enables researchers to perform complex multi-species plasma
simulations by combining reusable software components. This is
achieved by building on the BOUT++ framework of partial differential
equation solvers, and defining a flexible yet robust method of
coupling components together within a parallelised high-performance
code.

Applications of this tool to simulations of tokamak plasmas have been
demonstrated: Time dependent simulations of plasma filament/blob
propagation and steady-state transport including atomic
reactions. Convergence tests and comparisons to analytic solutions
have been carried out, demonstrating good conservation properties and
convergence of the methods. The public Git repository includes a suite
of unit, integrated and Method of Manufactured Solutions (MMS) tests
that are used routinely to check the correctness of code changes.

Areas for future development and research have been identified:
Extending the steady state solver implemented using PETSc from 1D
transport problems (section~\ref{sec:1d-transport}) to 2D is a high
priority, as is benchmarking of Hermes-3 against other codes for both
transport and turbulence applications. These efforts have begun and
will be reported elsewhere once completed.

Hermes-3 is publicly available~\cite{dudson:hermes3} on Github under a
GPL-3 license. To maximise its utility to the plasma community a set
of examples are included, and a manual~\cite{dudson:hermes3-manual}
provides an introduction for new users.

\section*{Acknowledgements}

This work was in part performed under the auspices of the U.S. DoE by
LLNL under Contract DE-AC52-07NA27344, and received funding from LLNL
LDRD 23-ERD-015. B.Dudson would like to thank Dr. Wayne Arter (CCFE)
for useful discussions and suggestions. The Hermes-3 source code,
simulation inputs, and processing scripts needed to reproduce the
results shown in this paper are available at
\url{https://github.com/bendudson/hermes-3}, Git commit
\texttt{5f56919} (Version 1.1.0). LLNL-CODE-845139.

\clearpage
\bibliographystyle{elsarticle-num}
\bibliography{bibliography}

\appendix

\newpage
\section{Numerical method for parallel dynamics}
\label{apx:numerics1d}

Dynamics parallel to the magnetic field are solved using a
2$^{nd}$-order slope-limiter method, briefly described in
section~\ref{sec:finite_differencing}.  For any number of fluids we
solve the number density $n$, momentum along the magnetic field,
$mnv_{||}$, and either pressure $p$ or energy $\mathcal{E}$. Here $m$
is the particle mass, so that $mn$ is the mass density. $v_{||}$ is
the component of the flow velocity in the direction of the magnetic
field, and is aligned with one of the mesh coordinate directions.  All
quantities are cell centered.

Cell edge values are by default reconstructed using a MinMod method
(other limiters are available, including 1st-order upwind, Monotonized
Central, and Superbee). If $f_i$ is the value of field $f$ at the
center of cell $i$, then using MinMod slope limiter the gradient $g$
inside the cell is:
\begin{equation}
g_i = \left\{\begin{array}{ll}
  0 & \textrm{if $\left(f_{i+1} - f_{i}\right) \left(f_{i} - f_{i-1}\right) < 0$} \\
  f_{i+1} - f_{i} & \textrm{if $\left|f_{i+1} - f_{i}\right| < \left|f_{i} - f_{i-1}\right|$} \\
  f_{i} - f_{i-1} & \textrm{Otherwise}
\end{array}\right.
\end{equation}
The values at the left and right of cell $i$ are:
\begin{align}
f_{i, R} &= f_i + g_i / 2 \nonumber \\
f_{i, L} &= f_i - g_i / 2
\end{align}
This same reconstruction is performed for $n$, $v_{||}$ and $p$ (or
$\mathcal{E}$).  The flux $\Gamma_{i+1/2}$ between cell $i$ and $i+1$
is:
\begin{equation}
\Gamma_{f, i+1/2} = \frac{1}{2}\left(f_{i,R} v_{||i,R} + f_{i+1,L}v_{||i+1,L}\right) + \frac{a_{max,i+1/2}}{2}\left(f_{i,R} - f_{i+1,L}\right)
\end{equation}
This includes a Lax flux term that penalises jumps across cell edges,
and depends on the maximum local wave speed, $a_{max}$. Momentum is
not reconstructed at cell edges; Instead the momentum flux is
calculated from the cell edge densities and velocities:
\begin{equation}
\Gamma_{nv, i+1/2} = \frac{1}{2}\left(n_{i,R} v_{||i,R}^2 + n_{i+1,L}v_{||i+1,L}^2\right) + \frac{a_{max,i+1/2}}{2}\left(n_{i,R}v_{||i,R} - n_{i+1,L}v_{||i+1,R}\right)
\end{equation}

The wave speeds, and so $a_{max}$, depend on the model being solved,
so can be customised to e.g include or exclude Alfv\'en waves or
electron thermal speed. For simple neutral fluid simulations it is:
\begin{equation}
a_{max, i+1/2} = \max\left(\left|v_{||i}\right|, \left|v_{||i+1}\right|, \sqrt{\frac{\gamma p_{i}}{mn_i}}, \sqrt{\frac{\gamma p_{i+1}}{mn_{i+1}}}\right)
\end{equation}

The divergence of the flux, and so the rate of change of $f$ in cell
$i$, depends on the cell area perpendicular to the flow, $A_i$, and cell volume $V_i$:
\begin{equation}
\nabla\cdot\left(\mathbf{b} f v_{||}\right)_{i} = \frac{1}{V_i}\left[\frac{A_{i} + A_{i+1}}{2}\Gamma_{f, i+1/2} - \frac{A_{i-1} + A_{i}}{2}\Gamma_{f, i-1/2}\right]
\end{equation}

\subsection{Boundaries}

At boundaries along the magnetic field the flow of particles and
energy are set by e.g.  Bohm sheath boundary conditions or no-flow
conditions. To ensure that the flux of particles is consistent with
the boundary condition imposed at cell boundaries, fluxes of density
$n$ and also $p$ or $\mathcal{E}$ are set to the simple mid-point
flux:
\begin{equation}
\Gamma_{f, i+1/2}^{boundary} = f_{i+1/2}v_{||i+1/2}
\end{equation}
where $f_{i+1/2} = \frac{1}{2}\left(f_{i} + f_{i+1}\right)$ and
$v_{||i+1/2} = \frac{1}{2}\left(v_{||i} + v_{||i+1}\right)$ are the
mid-point averages where boundary conditions are imposed.  It has been
found necessary to include dissipation in the momentum flux at the
boundary, to suppress numerical overshoots due to the narrow boundary
layers that can form:
\begin{equation}
\Gamma_{nv, i+1/2}^{boundary} = n_{i,R}v_{||i,R}v_{||i+1/2} + a_{max}\left[n_{i,R}v_{||i,R} - n_{i+1/2}v_{||i+1/2}\right]
\end{equation}
where $n_{i+1/2} = \frac{1}{2}\left(n_{i} + n_{i+1}\right)$.

\section{Plasma and neutral atom transport equations}
\label{sec:equations}

The equations solved in 1D in section~\ref{sec:1d-transport} and in 2D
in section~\ref{sec:applications-2d} are detailed here for
completeness.  A total of seven spatially varying quantities are
evolved in time: The deuterium ion and atom densities ($n_{d+}$ and
$n_d$); the flow of ions and atoms parallel to the magnetic field
($v_{||,d+}$ and $v_{||,d}$); and the pressure of the ions, atoms, and
electrons ($p_{d+}$, $p_{d}$ and $p_e$). SI units are used except
temperatures, which are in eV. In a 1D domain
(section~\ref{sec:1d-transport}) the anomalous diffusion terms are
omitted and all derivatives perpendicular to the magnetic field are
assumed to be zero.

The equations for the deuterium ion species density $n_{d+}$,
parallel velocity $v_{||,d+} \equiv \mathbf{b}\cdot\mathbf{v}_{d+}$ and
pressure $p_{d+} = en_{d+}T_{d+}$ are:
\begin{subequations}
\begin{align}
  \frac{\partial}{\partial t}n_{d+} =& -\nabla\cdot\left[\left(\mathbf{b}v_{||,d+} + \mathbf{v}_{\perp,d+}\right)n_{d+}\right] + R_{iz} - R_{rc} \\
  \frac{\partial}{\partial t}\left(m_{d+}n_{d+}v_{||,d+}\right) =& -\nabla\cdot\left[\left(\mathbf{b}v_{||,d+} + \mathbf{v}_{\perp,d+}\right)m_{d+}n_{d+}v_{||,d+}\right] -\mathbf{b}\cdot\nabla p_{d+} \nonumber \\
  & + en_{d+}E_{||} + R_{cx}m_d\left(v_{||,d} - v_{||,d+}\right) \nonumber \\
  & - R_{rc}m_dv_{||,d+} + R_{iz}m_dv_{||,d} \\
  \frac{\partial}{\partial t}\left(\frac{3}{2}p_{d+}\right) =& -\nabla\cdot\left[\left(\mathbf{b}v_{||,d+} + \mathbf{v}_{\perp,d+}\right)\frac{5}{2}p_{d+}\right] + v_{||,d+}\mathbf{b}\cdot\nabla p_{d+} \nonumber \\
  &+ \nabla\left(\mathbf{b}\kappa_{||,d+}\mathbf{b}\cdot\nabla T_{d+}\right) + \nabla\cdot\left(\chi_{d+}n_{d+}\nabla_\perp T_{d+}\right) \nonumber \\
  &+ \frac{1}{2}m_d \left(R_{cx} + R_{iz}\right) \left(v_{||,d} - v_{||,d+}\right)^2  \nonumber \\
  &+R_{iz}\frac{3}{2}eT_d - R_{rc}\frac{3}{2}eT_{d+} + W_{d+,e}
\end{align}
\end{subequations}
where $\mathbf{b} = \mathbf{B}/B$ is the unit vector in the direction
of the magnetic field, and the gradient in the plane perpendicular to
the magnetic field is $\nabla_\perp \equiv \nabla -
\mathbf{b}\mathbf{b}\cdot\nabla$. Particle diffusion across the
magnetic field is implemented as a cross-field ion drift velocity
$\mathbf{v}_{\perp,d+}$ with diffusion coefficient $D$:
\begin{equation}
  \mathbf{v}_{\perp,d+} = -D\frac{1}{n_{d+}}\nabla_\perp n_{d+}
\end{equation}
The charge exchange (CX), ionization (IZ) and recombination (RC)
reactions between species have rates (events per m$^3$ per second):
\begin{subequations}
\begin{align}
  R_{cx} =& n_{d+}n_d\left<\sigma v\right>_{cx} \\
  R_{iz} =& n_en_d\left<\sigma v\right>_{iz} \\
  R_{rc} =& n_en_{d+}\left<\sigma v\right>_{rc}
\end{align}
\end{subequations}
where the Maxwellian-averaged cross sections $\left<\sigma v\right>$
are taken from Amjuel~\cite{amjuel}, specifically Amjuel reaction H.4
2.1.5 (ionisation), H.4 2.1.8 (recombination) and H.3 3.1.8 (charge
exchange).  Hydrogenic charge-exchange reactions are adjusted for
isotope mass, ion and neutral temperatures by calculating an effective
temperature $T_{eff} = T_{atom} / A_{atom} + T_{ion} / A_{ion}$ as
described in the Amjuel manual.

There is a transfer of thermal energy to ions from electrons due to
collisions, $W_{d+,e}$:
\begin{equation}
W_{d+,e} = 3\nu_{d+,e}n_{d+}\frac{m_{d+}}{m_{d+} + m_e}e\left(T_e - T_{d+}\right)
\end{equation}
with ion-electron collision frequency $\nu_{d+,e} = \nu_{e,d+}m_e / m_{d+}$.

The electron density $n_e = n_{d+}$ is set by quasineutrality; the
electron parallel velocity $v_{||,e} = v_{||,d+}$ from assuming that
the parallel current is zero (Note that this is a choice in this
particular model, not a general feature of Hermes-3).  The electron
pressure equation is:
\begin{subequations}
  \begin{align}
    \frac{\partial}{\partial t}\left(\frac{3}{2}p_e\right) =& -\nabla\cdot\left[\left(\mathbf{b}v_{||,e} + \mathbf{v}_{\perp,d+}\right)\frac{5}{2}p_e\right] + v_{||,e}\mathbf{b}\cdot\nabla p_e \nonumber \\
    &+ \nabla\left(\mathbf{b}\kappa_{||,e}\mathbf{b}\cdot\nabla T_e\right) + \nabla\cdot\left(\chi_een_e\nabla_\perp T_e\right) \nonumber \\
    & - E_{iz} + E_{rc} - W_{d+,e}
  \end{align}
\end{subequations}
where $E_{iz}$ and $E_{rc}$ are the energy cost and gain due to
ionization and recombination atomic processes respectively. These are
calculated using Amjuel~\cite{amjuel}, reactions 2.1.5 and 2.1.8.
Ionization always removes energy from the electrons; Recombination may
be either a source or sink of electron energy, depending on the
temperature and density.  Electron force balance is used to calculate
the parallel electric field $E_{||} \equiv \mathbf{b}\cdot\mathbf{E}$
and so transfer electron pressure $p_e$ forces to the ions:
\begin{equation}
  en_eE_{||} = -\mathbf{b}\cdot\nabla p_e -\nabla\cdot\left[\mathbf{v}_{\perp,d+} m_en_ev_{||,e}\right]
\end{equation}

The equations for the neutral deuterium atom density $n_d$, parallel
velocity $v_{||,d}$ and pressure $p_d = en_dT_d$ are:
\begin{subequations}
  \begin{align}
    \frac{\partial}{\partial t}n_d =& -\nabla\cdot\left[\left(\mathbf{b}v_{||,d} + \mathbf{v}_{\perp,d}\right)n_d\right] - R_{iz} + R_{rc}  \\
    \frac{\partial}{\partial t}\left(m_dn_dv_{||,d}\right) =& -\nabla\cdot\left[\left(\mathbf{b}v_{||,d} + \mathbf{v}_{\perp,d}\right)m_dn_dv_{||,d}\right] -\mathbf{b}\cdot\nabla p_d \nonumber \\
    &- R_{cx}m_d\left(v_{||,d} - v_{||,d+}\right) + R_{rc}m_dv_{||,d+}\nonumber \\
    & - R_{iz}m_dv_{||,d} \\
    \frac{\partial}{\partial t}\left(\frac{3}{2}p_d\right) =& -\nabla\cdot\left[\left(\mathbf{b}v_{||,d} + \mathbf{v}_{\perp,d}\right)\frac{5}{2}p_d\right] + v_{||,d}\mathbf{b}\cdot\nabla p_d \nonumber \\
    &+ \nabla\left(\kappa_d \nabla T_d\right) + \frac{1}{2}m_d\left(R_{cx} + R_{rc}\right)\left(v_{||,d} - v_{||,d+}\right)^2 \nonumber \\
    &-R_{iz}\frac{3}{2}eT_d + R_{rc}\frac{3}{2}eT_{d+}
  \end{align}
\end{subequations}
The flow of neutral atoms across the magnetic field,
$\mathbf{v}_{\perp,d}$, is derived by balancing friction forces
against pressure gradient~\cite{rognlien-2002}:
\begin{equation}
  \mathbf{v}_{\perp,d} = - \frac{T_d}{m_d \nu_d p_d}\nabla_\perp p_d
\end{equation}

The thermal conduction coefficients for each species are:
\begin{equation}
  \kappa_{||,d+} = 3.9 \frac{p_{d+}}{m_{d+}\nu_{d+}} \qquad \kappa_{||,e} = 3.16 \frac{p_e}{m_e\nu_e} \qquad  \kappa_d = \frac{p_d}{m_d \nu_d}
\end{equation}
The collision frequencies for each species, $\nu_{d+}$, $\nu_e$ and $\nu_d$, are:
\begin{subequations}
  \begin{align}
    \nu_{d+} =& \nu_{d+,d+} + \frac{m_e}{m_{d+}}\nu_{e,d+} + n_d\left<\sigma v\right>_{cx} \\
    \nu_e =& \nu_{e,d+} + \nu_{e,e} \\
    \nu_d =& n_{d+}\left<\sigma v\right>_{cx} + n_da_0\sqrt{2eT_d/m_d}
  \end{align}
\end{subequations}
The collision frequency of charged species $a$ on charged species $b$
is given by~\cite{hinton1984}:
\begin{equation}
\nu_{a,b} = \frac{q_aq_b n_b \log\Lambda\left(1 + m_a/m_b\right)}{3\pi^{3/2}\epsilon_0^2m_a^2\left(v_a^2 + v_b^2\right)^{3/2}}
\end{equation}
with $v_a^2 = 2T_a/m_a$.
Neutral-neutral collisions assume a kinetic diameter of $2.8\times
10^{-10}$m (cross-section $a_0 = 2.5\times 10^{-19}$m$^2$), chosen
based on typical values for species considered ($2.89\times 10^{-10}$m
for H$_2$, $2.60\times 10^{-10}$m for He, $2.75\times 10^{-10}$m for
Ne~\cite{Wikipedia-kinetic-diameter}).

The Coulomb logarithm is different for electron-electron, ion-ion and
electron-ion species interactions, and is calculated using the NRL
formulary~\cite{Huba2013} (page 34). Converted to SI units with $T$ in
eV the Coulomb logarithms are:
\begin{subequations}
  \begin{align}
    \log \Lambda_{e,e} =& 30.4 - 0.5\log n_e + \frac{5}{4}\log T_e - \sqrt{\epsilon + \left(\log T_e - 2\right)^2/16} \\
    \log \Lambda_{e,i} =& \left\{\begin{array}{c c}
    31 - 0.5\log n_e + \log T_e & \textrm{if } T_i\frac{m_e}{m_i} < 10Z_i^2 < T_e \\
    30 - 0.5\log n_e - \log Z_i + 1.5\log T_e & \textrm{if } T_i \frac{m_e}{m_i} < T_e < 10Z_i^2 \\
    23 - 0.5\log n_i + 1.5 \log T_i - \log\left(Z_i^2 A_i\right) & \textrm{if } T_e < T_i m_e / m_i
    \end{array}\right. \label{eq:lambda_ei} \\
    \log \Lambda_{i,i} =& 29.91 - \log\left[\frac{Z_1 Z_2\left(A_1 + A_2\right)}{A_1 T_2 + A_2 T_1}\sqrt{n_1Z_1^2/T_1 + n_2 Z_2^2 / T_2}\right]
  \end{align}
\end{subequations}
Note that earlier versions of the NRL formulary had a factor of
$A_i^{-1}$ rather than $A_i$ in equation~\ref{eq:lambda_ei}.

\end{document}